\documentclass[sigconf]{acmart}

\usepackage[ruled,linesnumbered]{algorithm2e} % For algorithms
\usepackage{booktabs} % For formal tables
\usepackage{balance} % For balanced columns on the last page
\usepackage{graphicx}
\usepackage{url}
\usepackage{caption}
\usepackage{amssymb} % assumes amsmath package installed
\usepackage{bm}
\usepackage{subcaption}
\usepackage{pdfpages}
\usepackage{capt-of}
\usepackage{tabularx}
\usepackage{epsfig}
\usepackage{xcolor}
\usepackage{multirow}
\usepackage{nomencl}
\usepackage{amsmath}
\usepackage{siunitx} 
\usepackage{enumitem}

\usepackage[mathscr]{euscript}
\newcommand{\sysname}{\texttt{milliMap}\xspace}
\newcommand{\buildA}{\emph{A}\xspace} % Wolfson Building
\newcommand{\buildB}{\emph{B}\xspace} % Robert Hooke Building

\title[milliMap]{See Through Smoke: Robust Indoor Mapping with Low-cost mmWave Radar}

\author{
Chris Xiaoxuan Lu$^{1,2}$, Stefano Rosa$^{1}$, Peijun Zhao$^{1}$, Bing Wang$^{1}$, Changhao Chen$^{1}$,
}

\author{John A. Stankovic$^{3}$, Niki Trigoni$^{1}$, Andrew Markham$^{1}$}

\affiliation{%
  \institution{$^{1}$ University at Oxford, Oxford, England, United Kingdom}
  \institution{$^{2}$ University of Liverpool, Liverpool, England, United Kingdom}
  \institution{$^{3}$ University of Virginia, Charlottesville, Virginia, USA}
}



\renewcommand{\shortauthors}{C. Lu et al.}

\begin{abstract}

This paper presents the design, implementation and evaluation of \sysname, a single-chip millimetre wave (mmWave) radar based indoor mapping system targetted towards low-visibility environments to assist in emergency response. A unique feature of \sysname is that it only leverages a low-cost, off-the-shelf mmWave radar, but can reconstruct a dense grid map with accuracy comparable to lidar, as well as providing semantic annotations of objects on the map. \sysname makes two key technical contributions. First, it autonomously overcomes the sparsity and multi-path noise of mmWave signals by combining cross-modal supervision from a co-located lidar during training and the strong geometric priors of indoor spaces. Second, it takes the spectral response of mmWave reflections as features to robustly identify different types of objects e.g. doors, walls etc. Extensive experiments in different indoor environments show that \sysname can achieve a map reconstruction error less than $0.2$m and classify key semantics with an accuracy of $\sim 90\%$, whilst operating through dense smoke. 

\end{abstract}

\copyrightyear{2020} 
\acmYear{2020} 
\setcopyright{acmcopyright}
\acmConference[MobiSys '20]{The 18th Annual International Conference on Mobile Systems, Applications, and Services}{June 15--19, 2020}{Toronto, ON, Canada}
\acmBooktitle{The 18th Annual International Conference on Mobile Systems, Applications, and Services (MobiSys '20), June 15--19, 2020, Toronto, ON, Canada}
\acmPrice{15.00}
\acmDOI{10.1145/3386901.3388945}
\acmISBN{978-1-4503-7954-0/20/06}

\begin{document}

\maketitle

%%%%%%%%%%%%%%%%%%%%%%%%%%%%%%%%%%%%%%%%%%%%%%%%%%%%%%%%%%%%%%%%%%%%%%%%%%%%%%

%%%%%%%%%%%%%%%%%%%%%%%%%%%%%%%%%%%%%%%%%%%%%%%%%%%%%%%%%%%%%%%%%%%%%%%%%%%%%%%%

%!TEX root = ../main.tex
\section{Introduction}

% why a comprehensive map is important
Emergency responders are frequently exposed to harsh and dangerous environments, with consequent threat to life. Statistics collected by the Federal Emergency Management Agency \cite{fire_usa} report that over a 10-year period in USA, $2,775$ firefighters died on duty. Where there is a need to save and evacuate victims from a burning, collapsed or flooded building, it is vital for emergency responders to have increased situational awareness. In most search and rescue cases this requires, and begins with, making a map of the unknown environment \cite{dhekne2019trackio}. 
% When firefighters enter a burning building, they do it with little to no visibility. Firefighters are thus constrained to perform search-and-rescue operations following specific patterns that require them to constantly keep contact with the building's inner walls at all times.
Rather than relying entirely on firefighters to slowly explore the building, a promising alternative is to use mobile robots to rapidly survey and build the crucial map. Emergency personnel can then be re-localized accurately within the map and key features such as exit routes can be indicated.

%To prevent operational personnel entering unsafe regions, a promising way to obtain the crucial map is to let a mobile robot explore and map the environment first. The obtained map can then be accessed by an incident commander outside to take actions accordingly.

% limitation of existing sensors on mapping

State-of-the-art mapping sensors on mobile platforms (e.g., a smartphone or a mobile robot) use optical sensors, such as laser range scanners (lidar) \cite{surmann2003autonomous}, RGB cameras \cite{gao2014jigsaw,dong2015imoon} and stereo cameras \cite{henry2014rgb} to produce accurate indoor maps. However, not only are optical sensors impaired by the presence of airborne obscurants (e.g., dust, fog and smoke), their use cases are also significantly restricted by poor-illumination (e.g., dimness, darkness and glare). These adverse conditions regularly occur in emergency situations, e.g., dense smoke for firefighting. Acoustic sensor based mapping approaches, such as ultrasonic \cite{chong1999feature} and microphones \cite{pradhan2018smartphone,zhou2017batmapper}, are robust to lighting dynamics, but they either suffer from limited sensing range or become ineffective in noisy environments. 

%This unique capability makes it particularly useful in search and rescue scenarios, where teams of disposable robots explore in dark environments full of airborne particulates and then yield a reliable indoor map for first responders.
%  Moreover, owing to the adoption of beam-forming rather than mechanical rotation, single-chip mmWave radar solutions are physically small and light yet have a long sensing range.

% good property of mmWave, esp. the low-cost mmWave chips 
The demand of mapping in the above challenging situations motivates us to consider single-chip millimetre wave (mmWave) radar, which has recently emerged as an innovative low-cost, low-power sensor modality in the automotive industry \cite{timmwave}. A key advantage of mmWave radar is its imperviousness to adverse environmental conditions, such as smoke, fog and dust.  In the specific case of fire response, mmWave radars can `see' through smoke and help firefighters understand smoke-filled environments where many other optical sensors fail. Compared with the cumbersome lidar or mechanical radar (e.g., CTS350-X \cite{weston2018probably}), single-chip mmWave radars are lightweight and thus more able to fit payloads of micro robots and form factors of mobile or wearable devices.

% limitation of prior arts on using mmWave for mapping
Despite these advantages, mmWave-based mapping in indoor environments is still under-explored. The main issues lie in the strong indoor multi-path reflections as well as the sparse measurements returned by single chip radars. In extreme cases, we observe up to $75\%$ outliers due to multi-path reflections, along with more than two orders of magnitude lower point density than a lidar counterpart. 

\begin{figure*}[h]
\centering
  \includegraphics[width=0.92\linewidth]{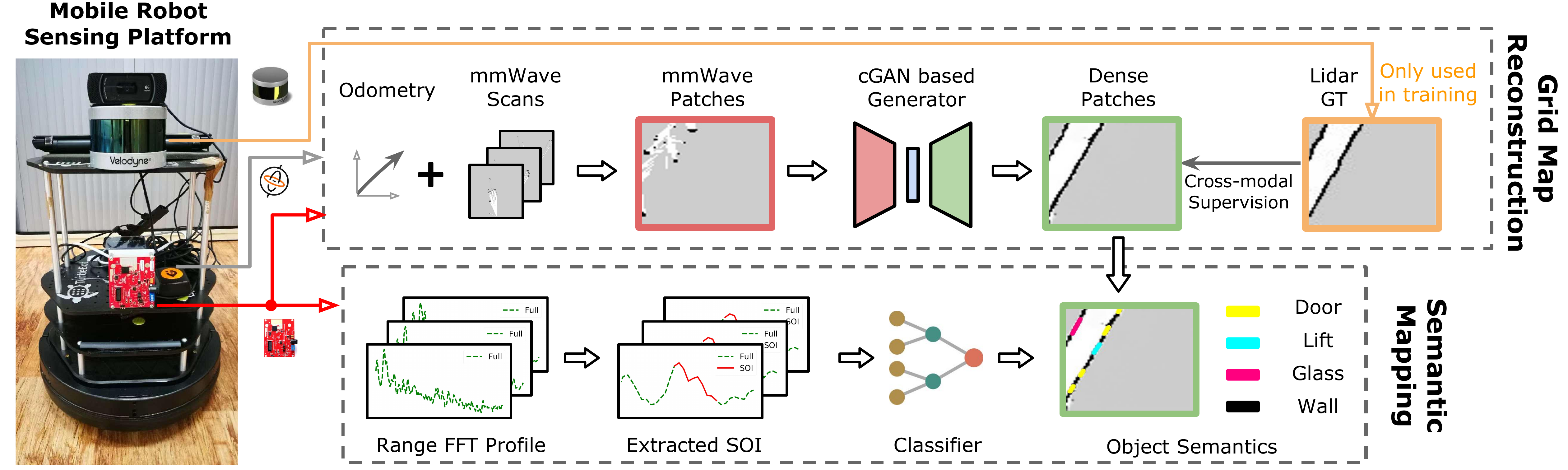}
  \caption{System overview of \sysname, comprising of (1) mobile robotic sensing (2) map reconstruction (3) semantic mapping.}
  \label{fig:system_overview}
\end{figure*}

To this extent, we propose \sysname, an approach overcoming the above issues to produce an occupancy grid map with semantic annotations on space accessibility, such as doors, lifts, glass, and walls. When taking emergency response into design consideration, a new set of design challenges arises.
\emph{First}, unlike \cite{wei2017facilitating} that aims to optimize mmWave network performance by pinpointing \emph{sparse} indoor reflectors with expensive SDRs, \sysname leverages a low-cost radar to reconstruct a \emph{dense map}. \emph{Second}, due to unknown floor plans and the demand of rapid response against disaster \cite{tashakkori2017indoor}, precisely moving a mmWave radar along pre-designed or navigated trajectories for object imaging is practically unfeasible, leaving prior solutions \cite{zhu2017object,zhu201560ghz} unsuitable in an emergency context. \emph{Third}, as building materials have complex internal layers and non-negligible diffusion effects \cite{kuga1996experimental,goulianos2017measurements}, previous identification methods only using the specular reflection from object surfaces \cite{zhu2015reusing} results in sub-optimal performance. 

\sysname tackles the above challenges via a novel mobile perception approach with the following contributions:
\begin{itemize}[leftmargin=*]
    \item A mobile robot based mapping system using single-chip mmWave radars for both occupancy grid mapping and semantic mapping in low-visibility indoor environments. 

    \item A generative learning approach that combines the cross-modal supervision from a co-located lidar and geometric priors of indoor spaces. Our approach overcomes the sparsity and noise issues of mmWave signals and is able to produce dense maps with an error less than $0.2$m.

    \item A semantic mapping method that robustly identifies objects by harnessing the multi-path effects of mmWave reflections, providing a classification accuracy $\sim 90\%$.

    \item A real-time prototype implementation with extensive real-world evaluations, including testing in smoke-filled conditions. 
\end{itemize}

The rest of the paper is organized as follows. We describe primer and system overview in Sec.~\ref{sec:primer} and Sec.~\ref{sec:overview} respectively. The proposed map reconstruction approach is introduced in Sec.~\ref{sec:grid_map_reconstruction}, followed by semantic mapping in Sec.~\ref{sec:semantic_mapping}. Sec.~\ref{sec:implementation} details our prototype implementation and we evaluate it in Sec.~\ref{sec:experiments}. We summarize related work in Sec.~\ref{sec:relatedwork} and limitations in Sec.~\ref{sec:limitations}, and conclude this work in Sec.~\ref{sec:conclusions}.

\section{Primer} % (fold)
\label{sec:primer}

\subsection{Principles of mmWave Radar} % (fold)
\label{sub:principles_of_mmwave_radar}

\noindent \textbf{Range Measurement}
The single chip mmWave radar uses a frequency modulated continuous wave (FMCW) approach \cite{uttam1985precision}, and has the ability to simultaneously measure both the range and relative radial speed of the target. In FMCW, a radar uses a linear `chirp' or swept frequency transmission. When receiving the signal reflected by an obstacle, the radar front-end performs a dechirp operation by mixing the received signal with the transmitted signals, which produces an Intermediate Frequency (IF) signal. Based on this IF signal, the distance $d$ between the object and the radar can be calculated as: 
\begin{equation}
d=\frac{f_{IF}c}{2S}
\end{equation}
where $c$ represents the light speed $3 \times 10^8m/s$, $f_{IF}$ is the frequency of the IF signal, and $S$ is the frequency slope of the chirp. In the presence of multiple obstacles at different ranges, a fast Fourier transform (FFT) is performed on the IF signal, where each peak after FFT represents one or more obstacles at a corresponding distance. 

\noindent \textbf{Angle Measurement}
A mmWave radar estimates the obstacle angle by using a linear receiver antenna array.
It works by emitting chirps with the same initial phase, and then simultaneous sampling from multiple receiver antennas. Based on the differences in phase of the received signals, the Angle of Arrival (AoA) for the reflected signal can be estimated \cite{rong2006angle}. Formally, the AoA estimated from any two receiver antennas can be calculated as:
\begin{equation}
\theta=sin^{-1}(\frac{\lambda \omega}{2\pi d})
\end{equation}
where $\omega$ denotes the phase difference, $d$ represents the distance between consecutive antennas and $\lambda$ is the wave length.
When multiple pairs of receiver antennas are available, sophisticated algorithms, such as beamforming \cite{haykin1993radar} and MUSIC \cite{odendaal1994two} can be used to obtain the AoA. At this point, the position of a reflecting obstacle can be jointly determined by AoA and ranging estimation.

\subsection{Generative Adversarial Networks} % (fold)
\label{sub:generative_adversarial_networks}

By extending deep neural networks (DNNs) to work in the generative context, Generative Adversarial Networks (GANs) \cite{goodfellow2014generative} trains two neural networks simultaneously: a generator $G$ and a discriminator $D$. A vanilla generator $G$ takes a noise vector as input and generates a data sample by evaluating $G$. When conditioned generation is needed, the noise vector can be replaced with an explicit source $s$, in which case $G$ becomes a conditional generator \cite{perarnau2016invertible}. The discriminator $D$, on the other hand is trained to distinguish between the real samples and the generated samples from $G$. Effectively, the discriminator provides feedback about the quality of the generated sample to $G$, which uses this feedback to generate better samples subsequently and combats the discriminator. Iteratively, the two neural networks play a competitive game and both become better at their respective tasks. As discussed later, we exploit this generative ability to create dense maps from sparse input.

% section background (end)

%!TEX root = ../main.tex

\section{\sysname Overview} % (fold)
\label{sec:overview}
We introduce \sysname, a mmWave radar based indoor mapping system to facilitate environment sensing and understanding under low-visibility conditions. \sysname takes as input the mmWave reflections from the environment captured by a low-cost, single-chip mmWave radar, and outputs a dense grid map with semantic annotation on obstacles. Fig.~\ref{fig:system_overview} shows the following modules in \sysname:

\noindent \textbf{Mobile Robot Sensing}. This module serves as the frontend, by which \sysname collects environment information from a mmWave radar and a lidar co-located on a mobile robot. Note the lidar is only used in the offline training phase to serve as ground truth/label provider. For online mapping phase, only the mmWave sensor is used.

\noindent\textbf{Grid Map Reconstruction}. Given the multi-modal data collection, this module uses a conditional GAN to reconstruct a dense grid map that depicts and marks obstacles, free spaces and unknown areas. In particular, this module features an autonomous learning fashion where our reconstruction model automatically leverages lidar samples as training supervision without human annotation. Once the training is over, the model can generate dense maps from mmWave signals alone, even in unseen low-visibility environments (e.g. smoke distribution) during training.

\noindent \textbf{Semantic Mapping}. The last module of \sysname is semantic mapping that classifies the obstacle semantics on the reconstructed grid map based mmWave reflection traits. Beyond simply using the specular reflections along direct paths, our recognizer considers and characterizes the multi-path effects to enhance the classification robustness.
% section overview (end)

%!TEX root = ../main.tex

\section{Grid Map Reconstruction} % (fold)
\label{sec:grid_map_reconstruction}

\begin{figure}[!t]
\centering
		  \includegraphics[width=\columnwidth]{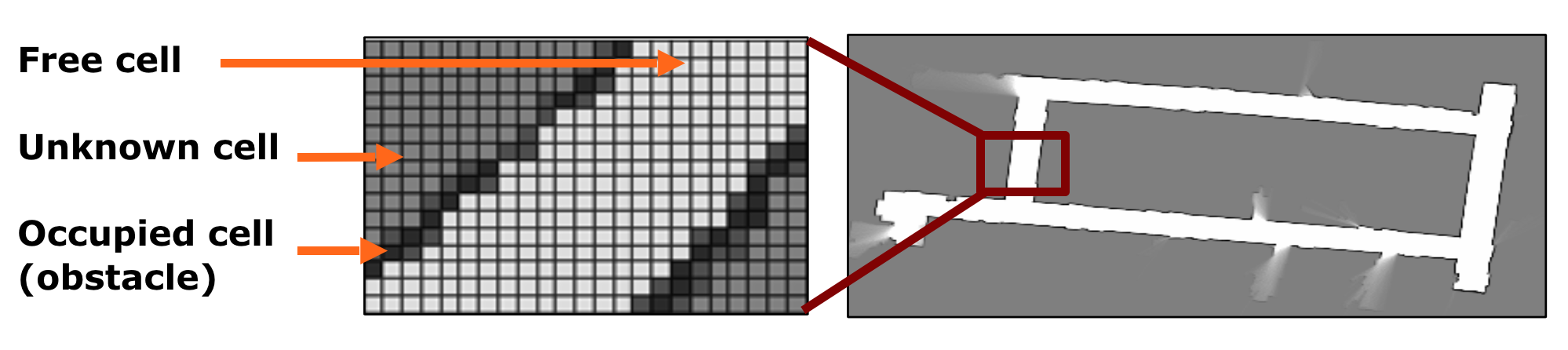}
  \caption{Bayesian grid mapping. Each cell in the map can represent free space (white), obstacles (black), or an unknown state (grey) if it has never been observed.}
  \label{fig:gridmapping}
 	\vspace{-0.5cm}
\end{figure}

The goal of map reconstruction is to generate a detailed and accurate map. In terms of map representation, this work uses an occupancy grid, which is widely used for mobile robot navigation \cite{thrun2005probabilistic} and can be easily understood by human users. As shown in Fig.~\ref{fig:gridmapping}, each cell (i.e., grid) on the map can be in one of three states: ``free" when it is empty, ``occupied" when it contains an obstacle or ``unknown" when it has never been observed. With these three states, place reachability can be inferred, allowing safe and fast navigation.

\subsection{Challenges: Sparsity and Noise Issues} % (fold)
\label{sub:challengs}

Before diving into the technical details, we first study the challenges of mmWave based grid mapping.
A mmWave radar detects ambient objects based on signal reflection. After several on-board pre-processing steps (e.g., interference mitigation), the range and orientation of reflecting points can be estimated and these points collectively form a \emph{point cloud} in the field of view. However, unlike the dense point clouds generated by lidars or depth cameras, the mmWave point cloud in indoor environment has two fundamental issues: i) multi-path noise and ii) sparsity.

\subsubsection{Multi-path Noise} % (fold)
\label{ssub:multi_path_noises}

\begin{figure}[t]
  \centering
	\begin{subfigure}[b]{0.14\textwidth}\centering
		\includegraphics[width=\columnwidth]{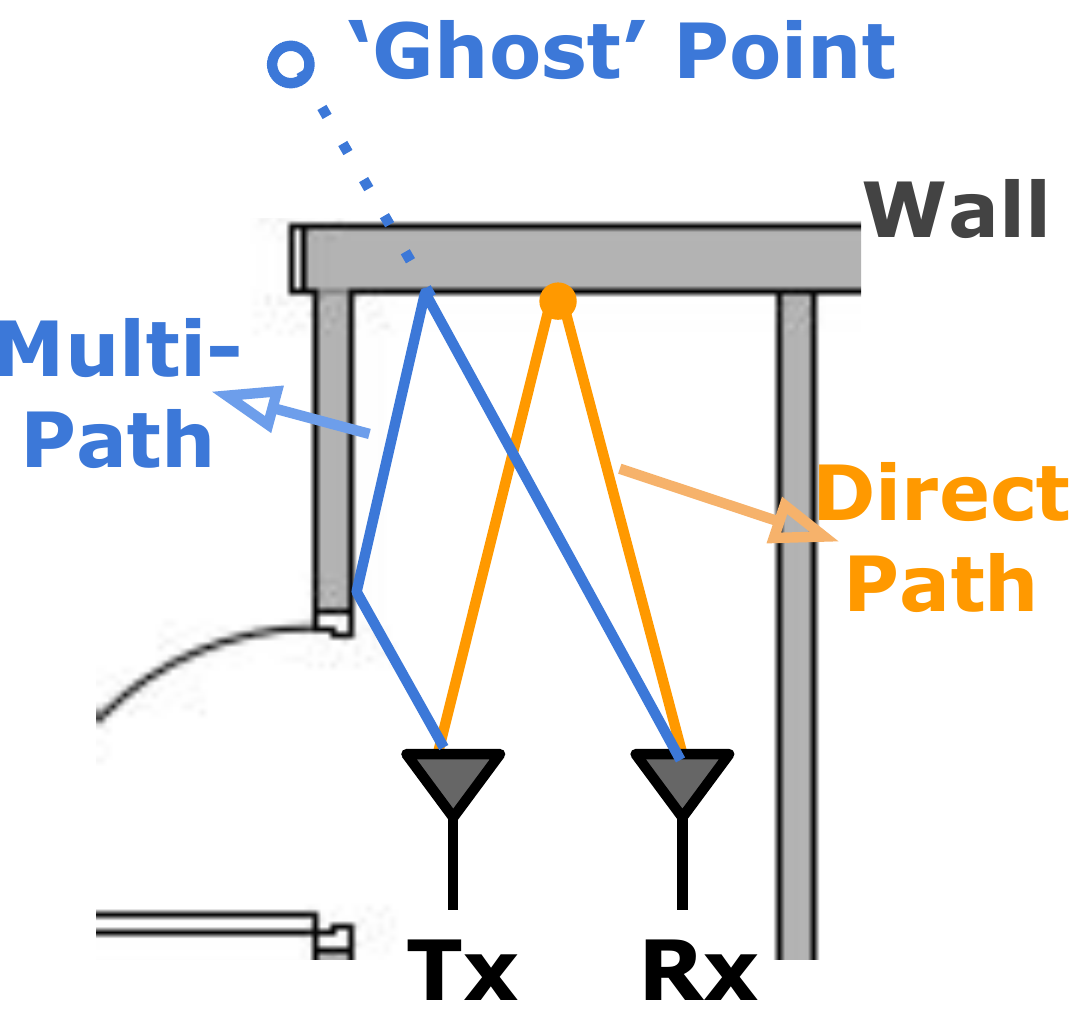} 
		\caption{Illustration.}
		\label{fig:illus}
	\end{subfigure}%
	\hfill
	\begin{subfigure}[b]{0.33\textwidth}\centering
		\includegraphics[width=\columnwidth]{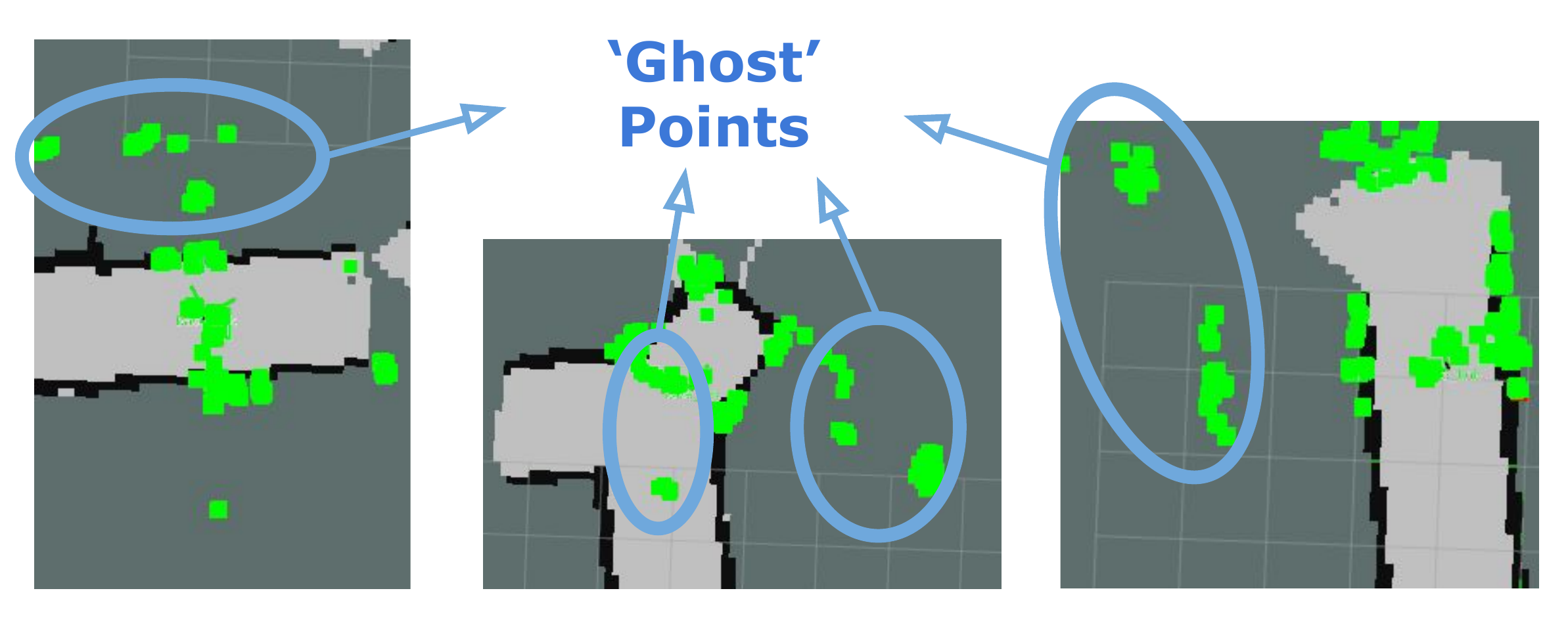} 
		\caption{Examples}
		\label{fig:ghost_points}
	\end{subfigure}%
	\caption{Multi-path Noise. The black lines in (\ref{fig:ghost_points}) are walls and there are non-negligible noise artefacts (in green) behind walls that are the result of multi-path reflection.}
	\label{fig:multi-path}
\end{figure}

Similar to any radio frequency technology, the signal propagation of MIMO mmWave in indoor environments is subject to multi-path issue due to aliasing from imperfect beams \cite{jog2019many} and reflection from surrounding objects (see Fig.~\ref{fig:illus}). As a consequence, reflected signals arriving at a receiver antenna are normally from two or more paths, leading to smearing and jitter. Multi-path is the primary contributor to the non-negligible proportion of pertinent noise artefacts or `ghost points' in a mmWave point cloud. Given $\sim 15$m bound of our indoor environment, we empirically found that, in extremely severe multi-path scenarios, e.g., corridor corners, ghost points can account for $>75\%$ points of a frame, which severely impacts grid mapping steps.
Fig.~\ref{fig:ghost_points} shows examples of noisy point clouds, where we can see many ghost points behind walls.

\vspace{-0.1cm}

\subsubsection{Sparsity} % (fold)
\begin{figure}[t]
\centering
  \includegraphics[width=0.9\linewidth]{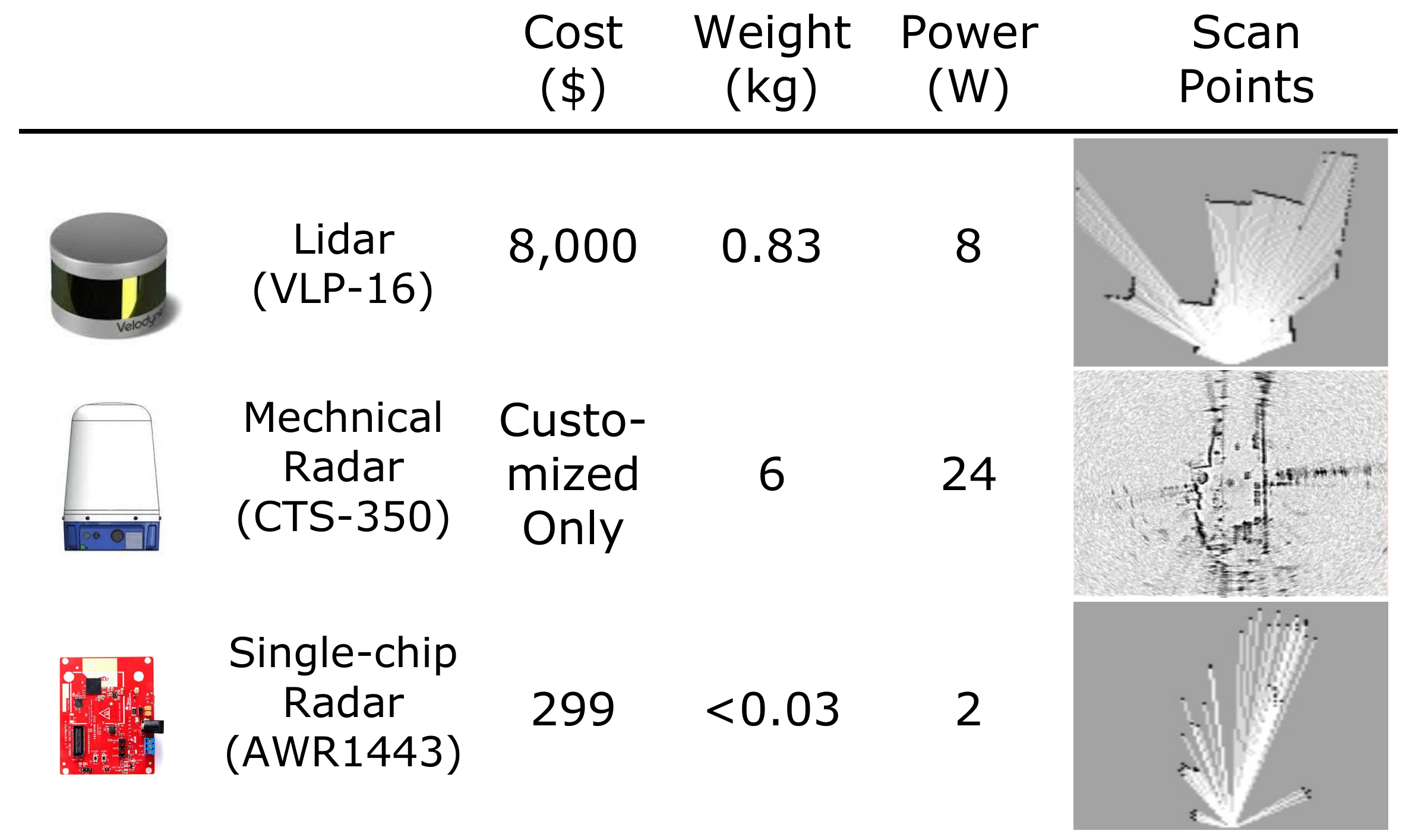}
  \caption{Comparison of lidar, mechanical radar and our single-chip radar. In each category, the features of a representative model are listed. Notably, compared with a lidar and a mechanical radar \cite{weston2018probably}, our beamforming radar is much cheaper and lighter, but only provides few points.}
  \label{fig:sensor_comparison}
 \vspace{-0.5cm}
\end{figure}

\label{ssub:sparsity}
%In addition, unlike a mechanically rotating/scanning radar, the beamforming radar used in this work is static with limited field of view.
As shown in Fig.~\ref{fig:sensor_comparison}, the point cloud given by a single-chip mmWave radar is approximately $\sim100$ reflective points per scan, which is over $100\times$ sparser than a lidar. 
Such sparsity results from three factors, including (1) the fundamental specularity of mmWave signals, (2) the low-cost single-chip design and (3) restricted sensing range by manually settings. Wireless mmWave signals are highly specular i.e., the signals exhibit mirror-like reflections from objects \cite{guan2019high}. As a result, not all reflections from the object propagate back to the mmWave receiver and major parts of the reflecting objects do not appear in the point cloud. Moreover, unlike massive array radar technology, due to cost and size constraints, the mmWave radar in our use only has $7$ antennas, which fundamentally limits its resolution. Moreover, as opposed to massive MIMO radar technologies, the mmWave radar in this case only has $3\times4$ antennas. Such a design is effective in both cost and size but results in poor angular resolution ($15^{\circ}$ in azimuth, $58^{\circ}$ in elevation) and targets which are closely spaced will be `smeared' together. 
Moreover, in order to lower bandwidth and improve signal-to-noise ratio, algorithms such as CFAR (Constant False Alarm Rate) \cite{ward1969handbook} are used for data processing and \emph{only} provide an aggregated point cloud, further reducing density. 
The third factor resulting in sparsity is specific to indoor mapping tasks and a consequence of multi-path noise. mmWave point clouds contain a non-negligible portion of `ghost points', which can mislead map densification. In order to suppress these `ghost points', we discard points outside of a sensing radius of $6$m, as multi-path effects generally incur false-positive points at longer distances \cite{yan2016multipath}. However, this restriction inevitably decreases the density of point clouds further.

% DEAD
%Specially, denoising can be easy when lots of redundant and complementary information is provided (e.g., channel information of RGB images); however this is beyond the reach of the sparse mmWave data. Densitification can also be relatively easy when the provided contextual information is precise and accurate (e.g., neighbor points of lidar outputs); unfortunately the multi-path noise in mmWave data is non-negligible that cannot give confident contexts.

\subsection{Reconstruction Framework} % (fold)
\label{sub:reconstruction_framework}

% why upsmapling and how difficult it is
With knowledge of the properties of mmWave data, \sysname aims to create a dense grid map. Owing to the complex interaction of the aforementioned challenges, this essentially requires an upsampling approach that can simultaneously address the sparsity and noise/outlier issues, which is far from trivial. 
Such a huge design challenge makes classic methods based on heuristics inadequate here (as seen in Sec.~\ref{sub:grid_map_reconstruction_performance}).

% Although traditional Inverse Sensor Models (ISM) techniques work well on high-fidelity sensors such as lidar, these ISM methods struggle to model challenging radar noises and often impose strict assumptions on the noise distribution \cite{weston2018probably}. In fact,  introduces huge challenges.  

\noindent \textbf{Reconstruction Neural Network.} To address the sparsity and noise challenge, we propose to use generative neural network (i.e., GAN in this work) reconstruct maps.
As discussed in Sec.~\ref{sub:generative_adversarial_networks}, conditional GAN is a learning paradigm that has proved to be a very effective tool for improving image resolution and generating realistic looking images. More importantly, GAN has the proven ability to \emph{reconstruct details} \cite{yang2018dense}, which can be crucial for route planning for search and rescue. Intuitively, GAN can utilize receptive fields in its CNN generator to denoise and densify image patches by referring to its neighboring contexts. Therefore, the generator in GAN can learn to fill in the missing gaps due to sparsity and eliminates artifacts caused by multipath. The discriminator in GAN further allows us to recover the underlying outline similar to the real ones. In fact, using GAN to perform denoising \cite{yan2017dcgans} and super resolution \cite{ledig2017photo} has become a predominant fashion in the computer vision field when heuristics fall short. 
Concretely, our adopted network architecture is constructed based on pix2pixHD \cite{wang2018high}, which is a recently proposed encoder-decoder framework based on conditional GAN \cite{mirza2014conditional}. It comprises of a generator $G$ and a discriminator $D$. 
In our context, the goal of the generator $G$ is to transform sparse and noisy patches to dense and clean images, while the discriminator $D$ aims to distinguish real images (i.e., partial environment maps) from the transformed ones.  As in many other generative networks, U-Net \cite{ronneberger2015u} is adopted as the backbone in our generator. To allow a large receptive field without large memory overhead, our network also uses multi-scale discriminators and downsamples the real and synthesized images by different factors to create an image pyramid of various scales. The discriminators are trained to distinguish real and generated images at various scales.

\noindent \textbf{Cross-modal Supervision by Collocation}: Training the above neural network requires a large number of labelled images. 
However in reality, actual maps are not always available and even when they are, maps can be outdated because in general most buildings do not precisely match with blueprints \cite{thrun2002probabilistic}. Manually calibrating each map incurs huge labor costs and is hard to scale. On the other hand, it is a common practice to use lidar to map indoor/outdoor environments \cite{zhang2014loam,tomoiagua2016indoor,kuka}. Modern lidar can be very accurate and we therefore consider to use it for creating a fresh map that is consistent with the mmWave radar observations.
To achieve such a generic and cheap labeling manner, \sysname adopts a cross-modal supervised learning fashion by using only partial labels (i.e., lidar patches) generated from a co-located lidar, allowing a robot to learn about the occupancy of the indoor environment by simply traversing an environment. After the learning phase, the mmWave radar on the robot is able to gain mapping skills from past experience and becomes capable of generating a lidar-like map \emph{independently}.

% \begin{figure}[t]
% \centering
%   \includegraphics[width=0.8\linewidth]{figures/millimap_overview.pdf}
%   \caption{Training the generator network with cross-modal supervision.}
%   \label{fig:millimap_overview}
% \end{figure}

\subsection{Network Input} % (fold)
\label{sub:network_input}

Given the above neural network, it is not immediately clear what representation of the inputs is best. Similar to most networks for image-to-image translation, our network expects image-like inputs, with a fixed, relatively low, number of channels and spatial correlations between neighbouring pixels. This is not met by the inherent irregularity of point clouds. We thus need to firstly convert the point cloud to an image-like representation and then use existing networks to process it. 

% the simplest scan representation would actually be a 1d vector with radial distance measurements

\noindent \textbf{Limitation of Scan Inputs.}
Perhaps the most straightforward representation is a virtual 2D laser \emph{scan} obtained from the 3D point cloud. 
After projecting each scan to a planar 2D image via raytracing, generative convolutional neural networks are able to take it as an input and generate a denser and denoised image. The dense images can then be converted back to angular distance measurements via raytracing and used for mapping. However, as the mmWave point cloud is very sparse, the converted scan image from each frame contains few spatial correlations between neighboring pixels. Directly feeding such non-informative images to a network incurs overfitting and hard to generalize in new environments \cite{theis2015note}. For these reasons as well as our goal for developing \emph{2D} maps (i.e., z-axis is not needed for end maps), in this work we chose to work directly on map \emph{2D} \emph{patches}.

% \noindent \textbf{Output Representation.}
% In particular, we use a log-odds representation to easily update the map cells based on new observations. Each map cell can be in one of three states: ``free" when it is empty, ``occupied" when it contains an obstacle, ``undecided" when it has never been observed. The probability $\hat{x}_i$ that map cell $x$ is in the ``occupied" state (e.g., it represents and obstacle) 
% given all measurements ${\mathcal{I}}_{0:t}$ is expressed using log-odds as:
% \begin{equation}
%     \mathcal{L}(\hat{x}_i \vert {\mathcal{I}}_{0:t}) =  \mathcal{L}(\hat{x}_i \vert {\mathcal{I}}_{0:t-1}) + \mathcal{L}(\mathcal{I}_t)
%     \label{eq:mapupdate}
% \end{equation}
% where:
% $$\mathcal{L}(\hat{x}_i)= log\Bigg[{{p(\hat{x}_i)} \over { 1- p(\hat{x}_i)}}\Bigg]$$. 

% $p(\hat{x}_i)$ is a prior on the probability of $x$ to be occupied. $\mathcal{L}$ is kept bounded between two limits $l_{min}$ and $l_{max}$.
% As with traditional grid mapping, \ref{eq:mapupdate} ensures that false classifications do not corrupt the map, since the probabilities within the cells are adapted gracefully and will be corrected by later observations. 
% A threshold of 0.7 is then applied to the cell probabilities to discretize the values of observed cells into free or occupied.
% Figure \ref{fig:gridmapping} shows a depiction of the grid mapping procedure.

\noindent \textbf{Patches as Input}
The way map patches are generated differs between the training and prediction phases.
During training, since we have access to the full, yet sparse, grid maps through running off-the-shelf Bayesian grid mapping \cite{octomap13}, we can generate patches by dividing the full map into a regular grid of patches of a given size ($6 \times 6m^2$ in this work), with an overlap of $50\%$.
However, at prediction time, we only generate patches along the robot's trajectory, in order to reduce inference time.
In particular, since we have access to a reasonably accurate odometry (e.g. from wheel odometry and/or inertial measurements), we can detect when the robot is moving out of the current patch, and extract a new patch along the direction of travel, without overlapping with the previous patch ($6 \times 6m^2$). This simplification ensures we don't have to merge two overlapping predictions.
We then feed patches of the generated map along with the past robot trajectory to our network for denoising and densification. 
The advantage of this hybrid approach is that patches are built in real-time, whilst the more expensive map densification process is only triggered when entering a new patch. 
Hereafter, we denote the reconstructed map patches as $\mathbf{x}$ and the noisy mmWave patches as $\mathbf{s}$. The pivotal goal of \sysname is to translate mmWave patches to dense map patches through a deep neural network. The dense patches are then stitched together to produce a full map.

\subsection{Reconstruction Loss Functions} % (fold)
\label{sub:objective}

The objective function of our network is comprised of losses from four sources: (1) a conditional GAN, (2) an intermediate feature matching, (3) a perceptual loss, and (4) a map prior. 

% In particular, the \emph{map-prior loss} is our proposed term that enforces indoor geometric consistency in the generated patches. 

\noindent \textbf{Reconstruction Likelihood.} We use conditional GANs to model the conditional distribution of real map patches $\mathbf{x}$ given the input mmWave map patches $\mathbf{s}$, which are converted from the sparse point cloud. The conditional GAN loss can be expressed as:
	\begin{equation*}
		\label{eq: cgan_loss}
		\begin{split}
        \mathcal{L}_{cGAN}(G, D_k) = &\mathbb{E}_{(\mathbf{s},\mathbf{x})} [\log D_k(\mathbf{s},\mathbf{x})] \\
        + &\mathbb{E}_{\mathbf{s}} [\log (1 - D_k(\mathbf{s},G(\mathbf{s}))]
        \end{split}
	\end{equation*}
where $G$ tries to minimize this objective function against an adversary network $D_k$ that tries to maximize it \cite{mirza2014conditional}. In particular, as our network uses multi-scale discriminators, $D_k$ here is the specific discriminator for $k$-th scale. 
In the meantime, to stabilize training and generate meaningful statistics at multiple scales, we follow \cite{dosovitskiy2016generating,wang2018high} and introduce the feature matching loss $\mathcal{L}_{FM}(G, D_k)$ in our objective function:
 	\begin{equation*}
		\label{eq: fm_loss}
        \mathcal{L}_{FM}(G, D_k) = \mathbb{E}_{(\mathbf{s},\mathbf{x})} \sum_{i=1}^{T} \frac{1}{N_i} || D^{(i)}_{k}(\mathbf{s},\mathbf{x}) - D^{(i)}_{k}(\mathbf{s}, G(\mathbf{s}))||_{1}
	\end{equation*}
where $T$ is the total number of layers, $D^{(i)}_{k}$ produces the features of $i$-th layer and $N_i$ denotes the number of nodes in that layer. \sysname computes this feature matching loss on multiple discriminators which is in line with our multi-scale architecture. Lastly, to compare high level differences and stabilize GAN training \cite{johnson2016perceptual}, we also introduce a perceptual loss in the objective function: 
 	\begin{equation*}
		\label{eq: perception_loss}
        \mathcal{L}_{VGG}(G) = \mathbb{E}_{(\mathbf{s},\mathbf{x})} \sum_{j=1}^{J} || F^{(j)}(G(\mathbf{s})) - F^{(j)}(\mathbf{x})||_{1}
	\end{equation*}
where $F$ is a pre-trained loss network used for image classification that helps to quantify the perceptual differences of the content between images. In this work, we follow \cite{johnson2016perceptual} and adopt the VGG network as $F$. Each layer $j$ in the VGG network measures different levels of perception.

\begin{figure}[!t]
\centering
		  \includegraphics[width=0.95\linewidth]{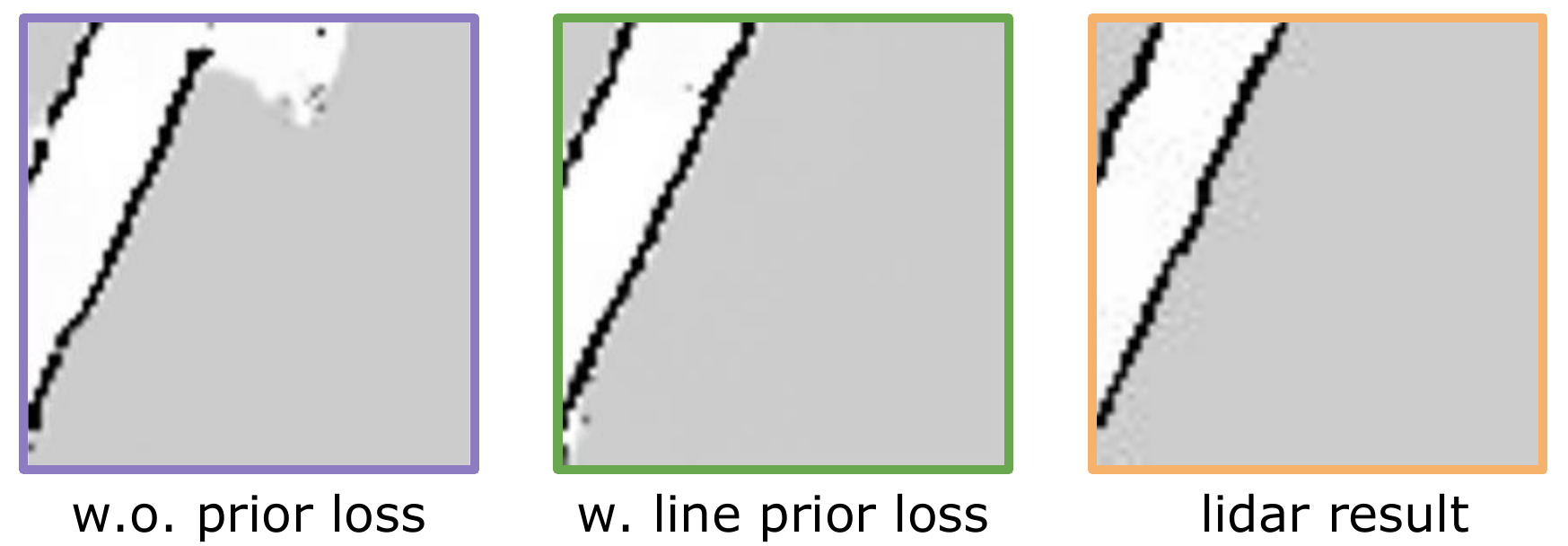}
  \caption{Effectiveness of map prior loss on a straight corridor patch. A line detector is used in this case to construct the map-prior loss and the produced `corridor' is straighter and more complete. lidar is used as pseudo-ground truth.}
  \label{fig:line_kernels}
\end{figure}

\noindent \textbf{Map Prior.}
The above losses only consider the efficacy of reconstruction in the latent space of high-level appearance but ignore the important low-level geometrics. Recent research found that
the latent spaces of appearance and geometry are not strongly correlated. Standard neural network generators can learn appearance transformation, however, lack the ability to embed complex geometry cues for effective image-to-image translation \cite{gokaslan2018improving,zhu2017unpaired}. Nevertheless, 2D indoor maps in modern buildings often have strong geometric structures that follow certain patterns, e.g. following rectilinear outlines for ease of construction. As this geometric information is fairly ubiquitous \cite{garulli2005mobile}, one can leverage it as a prior to bootstrap the patch generation process and enhance the quality of the final stitched map. Formally, given a generated patch $G(\mathbf{s})$ and its corresponding real patch $\mathbf{x}$, we define a \emph{map-prior loss} as follows:
 	\begin{equation}
		\label{eq: mp_loss}
        \mathcal{L}_{MP}(G) = \mathbb{E}_{(\mathbf{s},\mathbf{x})} \sum_{j=1}^{M} || \mathbf{h}^{(j)} \ast G(\mathbf{s}) - \mathbf{h}^{(j)} \ast \mathbf{x}||_{1}
	\end{equation}
where $\ast$ represents the convolution operator and $\mathbf{h}^{(j)}$ is one of $M$ convolution kernels with \emph{fixed} weights, determined by the types of convolution. For example, $\mathbf{h}^{(j)}$ can be a line or edge detection mask, capturing different geometric properties of images. Through a detector mask, this map-prior loss encourages the consistency between source and target patches corresponding to a certain geometric prior.  
For example, many objects (e.g., walls and doors) on indoor floor plans are line based \cite{garulli2005mobile}. Therefore, when using line detectors to embed such a prior in the loss, we can achieve better reconstruction performances in corridors, as shown in Fig.~\ref{fig:line_kernels}. Choices of convolution masks are flexible, mainly depending on the noise level of inputs as well as a particular map/building type. We will quantitatively discuss the impacts of different types of detectors in Sec.~\ref{sub:effectiveness_analysis_of_components}.

Finally, our full objective combines reconstruction likelihood and map prior as:
 	\begin{equation}
		\label{eq: total_loss}
		\begin{split}
        \mathcal{L}_{total} = \sum_{k=1,2, \ldots, K} &\mathcal{L}_{cGAN}(G, D_k) + \lambda_1 \mathcal{L}_{FM}(G, D_k) \\
        +& \lambda_2 \mathcal{L}_{VGG}(G) + \lambda_3 \mathcal{L}_{MP}(G)
        \end{split}
	\end{equation}
where $\lambda_1$, $\lambda_2$ and $\lambda_3$ are hyper-parameters for regularization. $K$ denotes the number of distinct scales for discriminators.

%!TEX root = ../main.tex

\section{Semantic Mapping} % (fold)
\label{sec:semantic_mapping}

% A comprehensive gridmap with both metric and semantic information will allow indoor robots to provide better context-aware services and robustly manoeuvre in clutter environments.

So far we have introduced how \sysname reconstructs a dense grid map from mmWave signals. Nevertheless, in order to best assist the decision making of emergency response, a thorough map should not only tell \emph{where} the obstacles are but also their \emph{semantics}. Exhausting the whole universe of indoor semantics is beyond the scope of this work; instead \sysname follows \cite{tashakkori2016facilitating} and focuses on $4$ predominant construction objects that semantically describe space accessibility: (1) horizontal access object (AO) - doors, (2) vertical AO - lifts, (3) alternative AO - glass and (4) non-AO - walls. 

\subsection{Complex Construction Objects} % (fold)
\label{sub:complex_construction_objects}

\begin{figure}[!t]
	\centering
	\begin{subfigure}[b]{0.23\textwidth}\centering
		\includegraphics[width=\columnwidth]{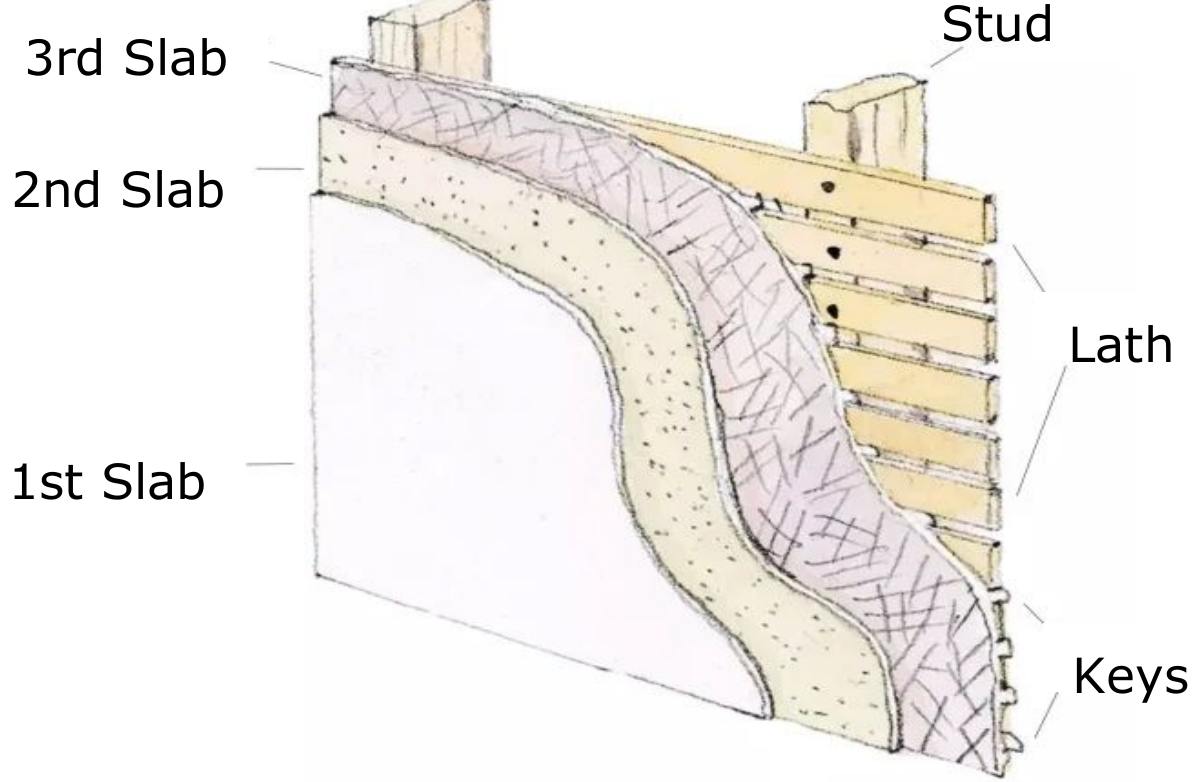} 
		\caption{}
		\label{fig:diagram_wall}
	\end{subfigure}%
	\begin{subfigure}[b]{0.25\textwidth}\centering
		\includegraphics[width=\columnwidth]{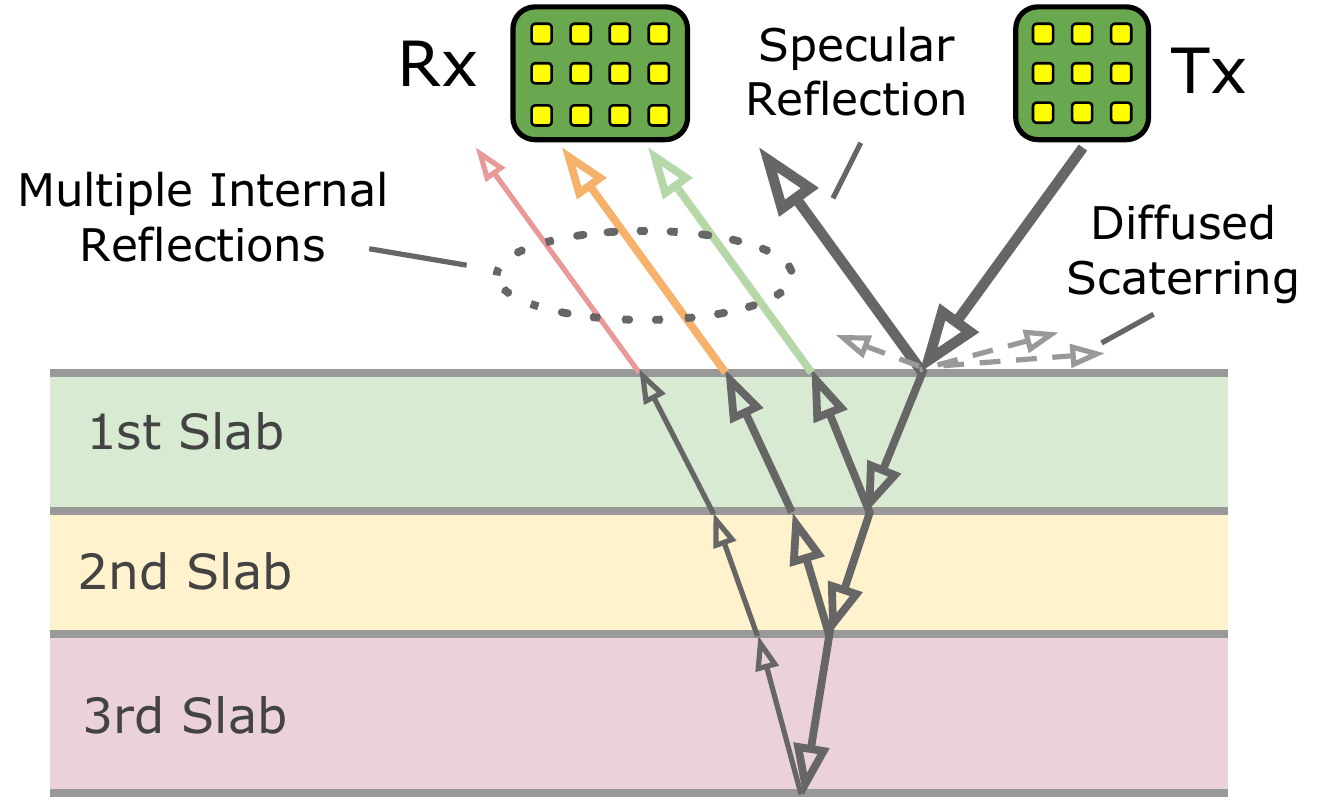} 
		\caption{}
		\label{fig:wall_reflections}
	\end{subfigure}%
\caption{mmWave signal propagation on a wall. (\ref{fig:diagram_wall}) A common interior building wall has multiple layers. (\ref{fig:wall_reflections}) The diffusion and multiple internal reflections on a simplified wall model (with only three slabs), result in complicated multi-path effects. We exploit these signatures for classification.}
\label{fig:wall}
\end{figure}

\noindent \textbf{Challenge.}
The main challenge here lies in the complexity of interior construction objects, with prior art on material identification difficult to directly apply. 
Specifically, previous work focuses on objects made of a single material or containing very thin layers (e.g., cardboard box). For these simple objects, the received mmWave signals are from the specular reflection from the object \emph{surface} and thus prior work (e.g., \cite{zhu2015reusing}) can directly use the \emph{strongest/peak} signal strength (RSS) value to determine the object type. However in our case, many construction objects in indoor environments, ranging from composite walls to hollow doors, consist of multiple slabs made from different materials. For instance, fig.~\ref{fig:diagram_wall} shows the diagram of common interior building wall, in which $5$ different layers are stacked together. Each of the slabs often has sufficient thickness that affects propagation characteristics of mmWave signals as well as resulting in multiple reflections from internal layers \cite{holloway1997analysis}. Additionally as discussed in \cite{kuga1996experimental,goulianos2017measurements}, building materials have different roughness and the diffusion effect of mmWave on some rough surfaces (e.g., the surface of wall) can be significant. Such diffusion effects, unfortunately, further complicates the problem of object identification (see Fig.~\ref{fig:wall_reflections}). Intuitively, the compound effect of diffusion, multiple internal reflections and specular reflection is hard to model by only using a peak RSS value.

\begin{figure*}[!ht]
	\centering
	\begin{subfigure}[b]{0.24\textwidth}\centering
		\includegraphics[width=\columnwidth]{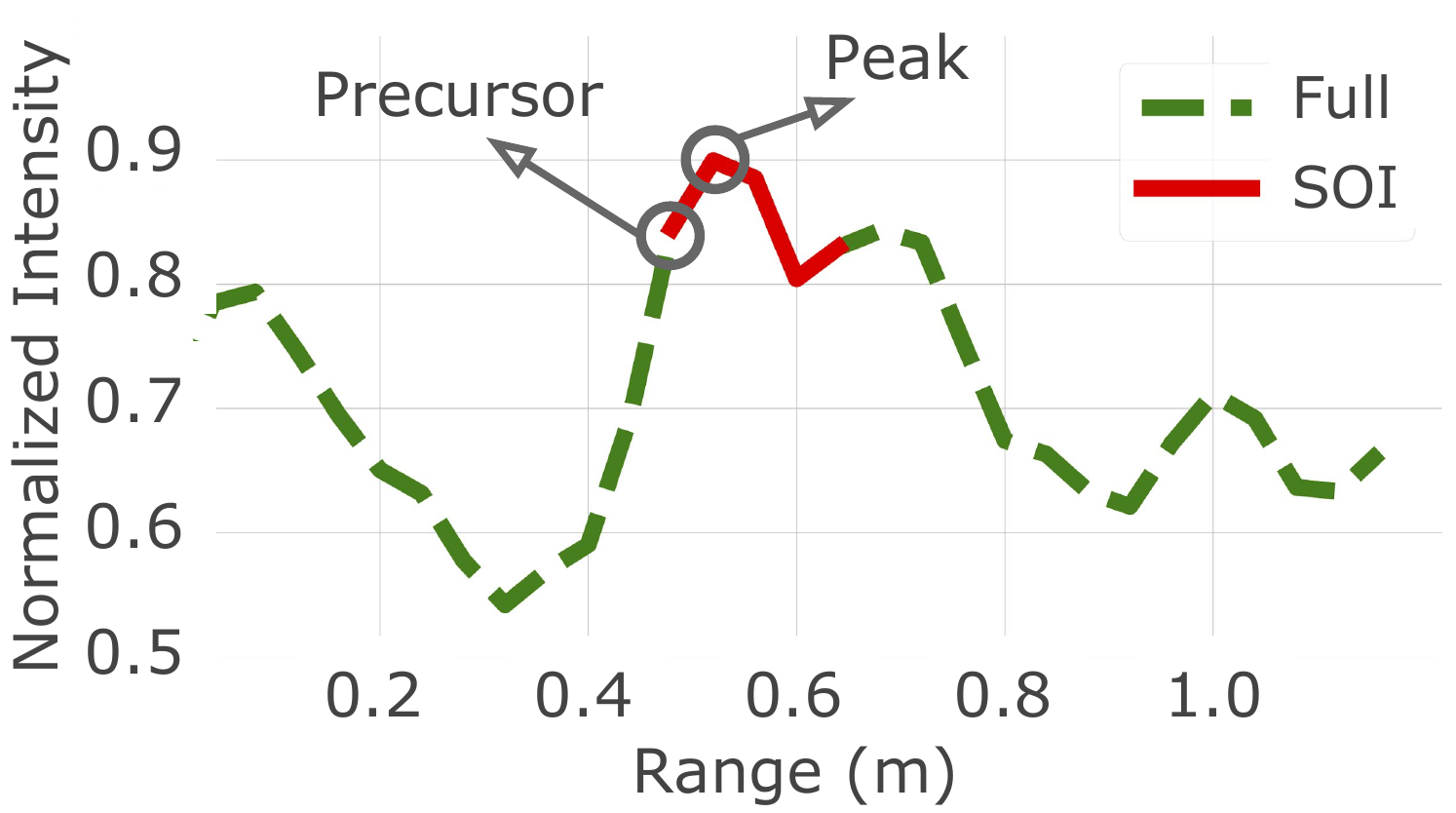} 
		\caption{SOI Extraction}
		\label{fig:soi_extraction}
	\end{subfigure}%
	\hfill
	\begin{subfigure}[b]{0.24\textwidth}\centering
		\includegraphics[width=\columnwidth]{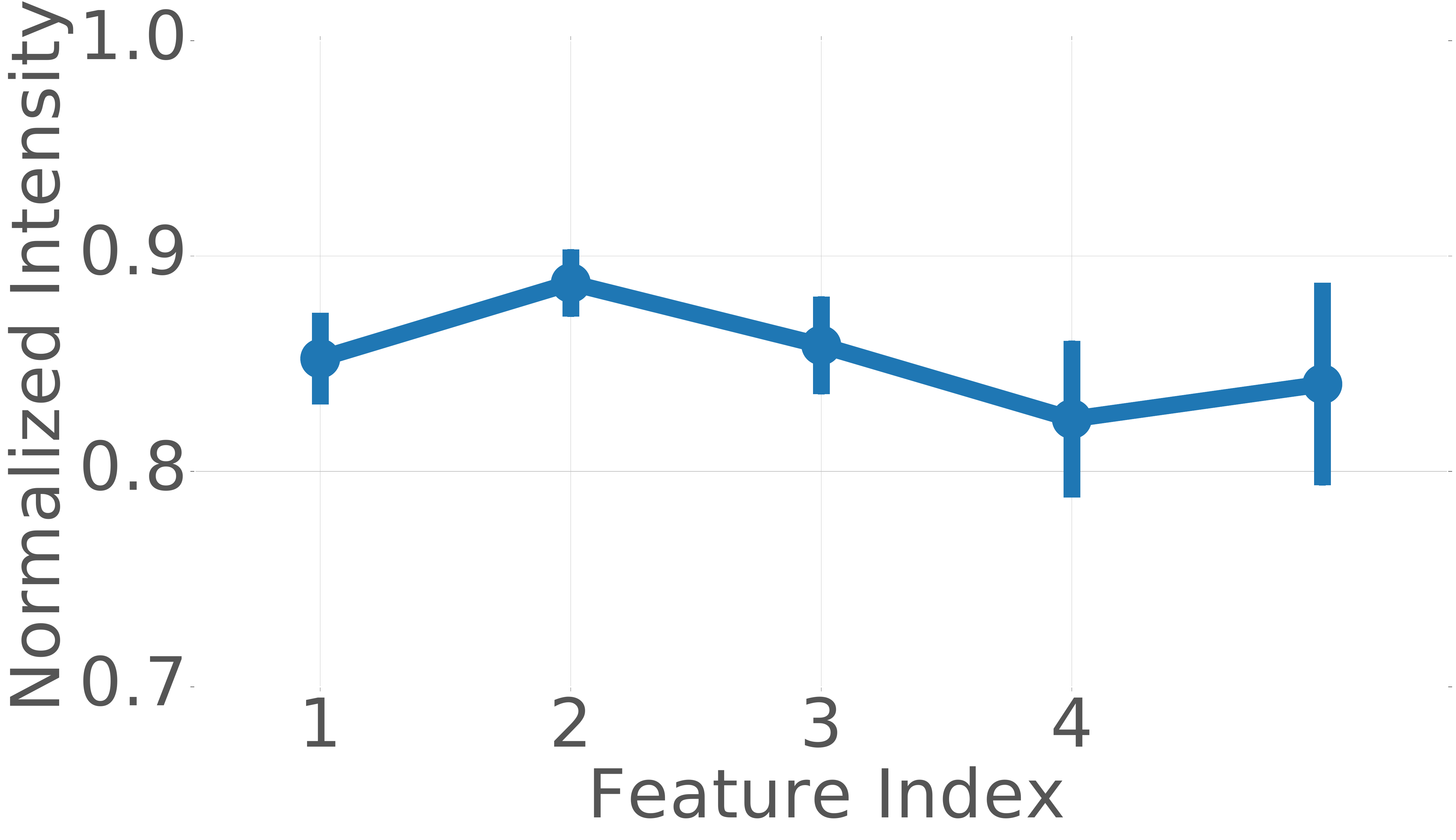} 
		\caption{Door}
		\label{fig:soi_door}
	\end{subfigure}%
	\hfill
	\begin{subfigure}[b]{0.24\textwidth}\centering
		\includegraphics[width=\columnwidth]{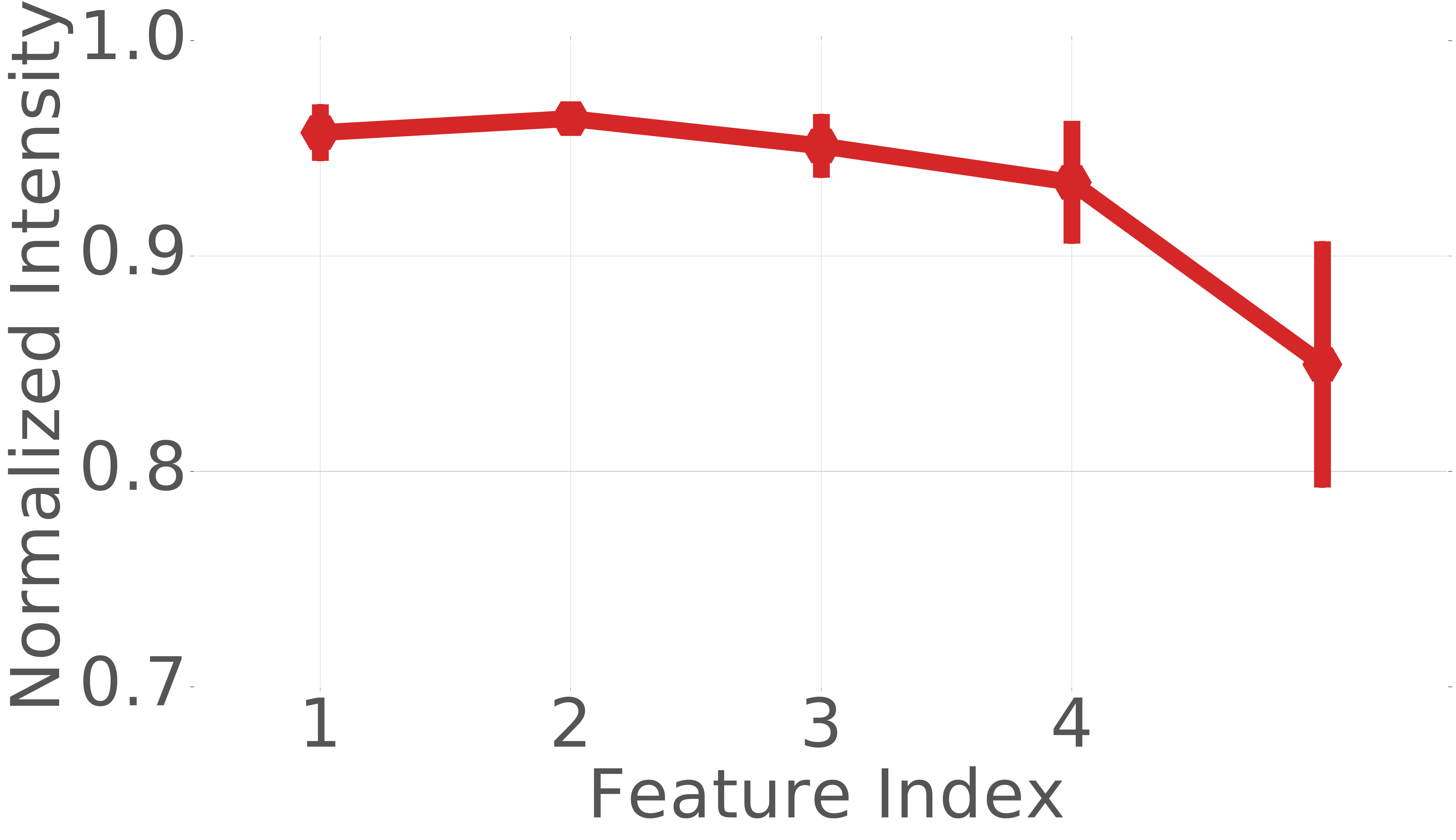} 
		\caption{Lift}
		\label{fig:soi_lift}
	\end{subfigure}%
	\hfill
	\begin{subfigure}[b]{0.24\textwidth}\centering
		\includegraphics[width=\columnwidth]{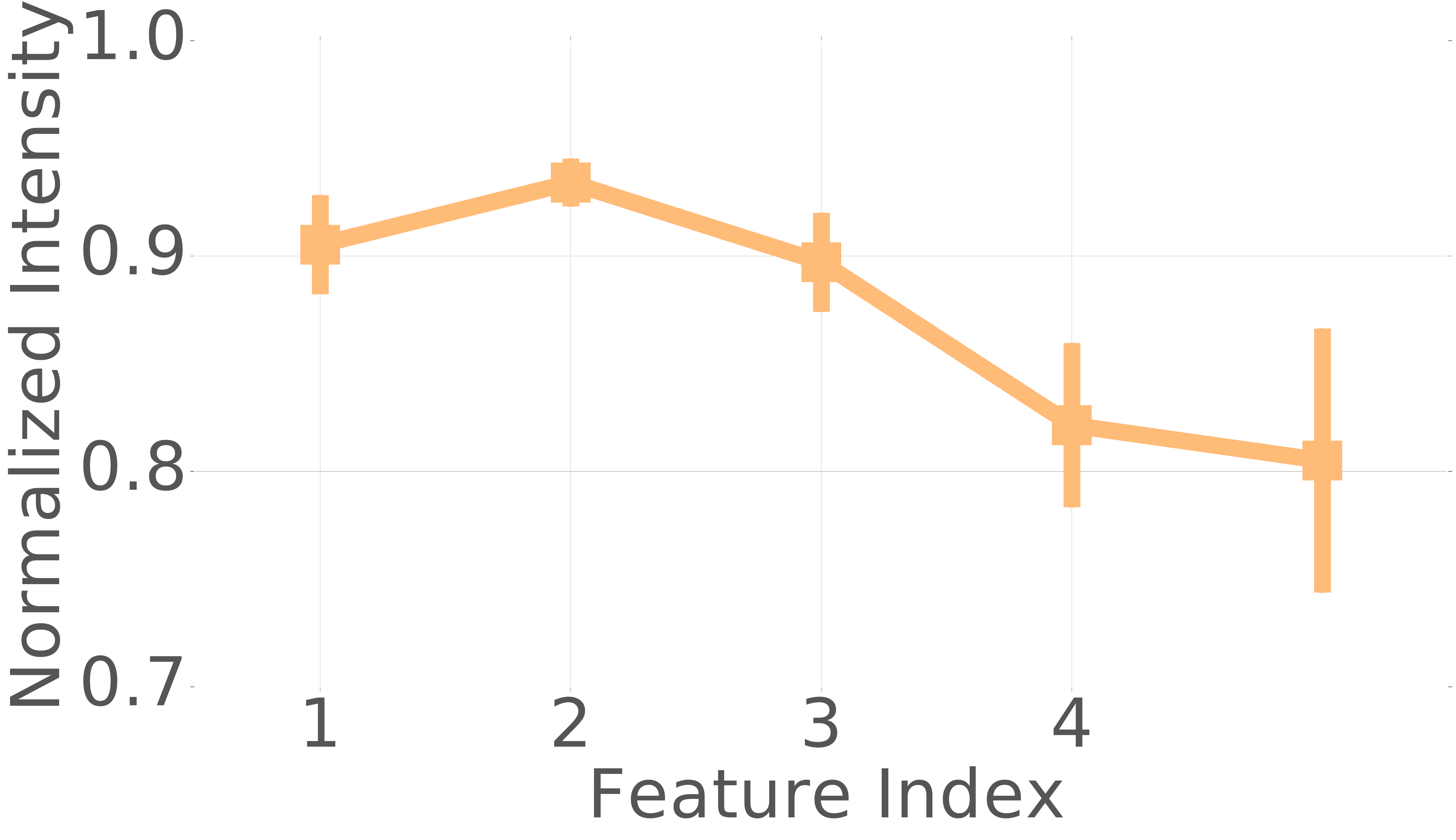} 
		\caption{Glass}
		\label{fig:soi_glass}
	\end{subfigure}%
\caption{Semantic Mapping: (a) $16$cm-wide SOI, determined by the corresponding peak in the range FFT profile; (b-d) `average' SOI aggregated from $27,952$ training samples. SOIs of different materials have distinct patterns. Note that the first feature index, namely the starting point in (b-d) is the precursor index to the detected peak value.}
\label{fig:soi_illustration}
\end{figure*}

\noindent \textbf{Key Idea and Observations.}
From the perspective of a receiver, both diffusion and multiple internal reflections cause multi-path effects. Owing to differences in several properties, such as roughness and interior layers, the multi-path effects exhibit certain patterns, captured in the 1D range FFT profile (see Sec.~\ref{sub:principles_of_mmwave_radar} for definition). Fig.~\ref{fig:soi_extraction} shows an example of a range FFT profile. The peak value in this example represents the normalized intensity of the specular reflection along the direct path, where neighbor values around it are due to multi-path effects from diffusion and multi-reflections. To illustrate what patterns we can extract from the shape of the peak, we extract features (e.g. peak value, standard deviation) from $27,952$ collected profiles of $3$ common construction objects. Fig.~\ref{fig:soi_door}, \ref{fig:soi_glass} and \ref{fig:soi_lift} show the average value and standard deviations, from which two key observations can be drawn. First, peak value differences (feature index $2$) between construction objects can be vague (e.g., glass versus lift) that confuses object classification. Second, both the magnitude and shape of neighboring points exhibit more distinct patterns, providing better object signatures.

\subsection{Semantic Recognizer} % (fold)
\label{sub:semantic_recognizer}

Based on the above observations, we propose a semantic recognizer that operates by first extracting a segment of interest from the range FFT profile, and then using a classifier to identify different types of obstacles.

\noindent \textbf{Segment of Interest.}
Notably, the first step before performing segment extraction is to acquire a scan at a perpendicular angle to the object. To combat the limited angular resolution of the TI board (see Sec.~\ref{sec:primer}, \sysname tasks the robot platform to mechanically scan its horizontal field of view, and then determines the perpendicular angle by pinpointing the pose that yields the largest reflection intensity. Once a perpendicular pose is determined, the robot platform enters the static mode and starts to record the range profile at this instant.
A practical issue of applying the above intuition is determining the number of points to consider after the peak, namely finding a \emph{segment of interest (SOI)} in the range profile. As multiple objects are in the mmWave radar's field of view, a range profile often contains extraneous information corresponding to non-target objects. Directly using the whole profile as features will thus confuse a \emph{single} object classifier. As the target object in our case is the nearest object perpendicular to the robot/radar, the starting point of a SOI is easy to find because it has the steepest increasing gradient in the profile. To mitigate the potential aliasing issue due to $40$mm ranging resolution, we always use the prior index to the steepest point as the starting point of SOI.
We empirically found that at a SOI width of $6$ points, the best tradeoff can be achieved. In Sec.~\ref{sub:semantic_mapping_performance}, we will further discuss the impact of different SOI widths on semantic classification. Fig.~\ref{fig:soi_extraction} illustrates the SOI extraction process.

% the maximum value in practice does not necessarily correspond to . The position of steepest increasing gradient rather than the max out approach allows robust detection of the starting point. 

% The endpoint of a SOI, is more sophisticated to determine that needs to resort to magnitude variation. In principle, the magnitude variation of a point in the range profile should be small if it corresponds to the diffusion and/or multi-reflection effects, which are intrinsic properties of objects. In contrast, the variation is large when the corresponding points are clutter or noise. 

% Interestingly, although the target object is the closest object perpendicular to the robot, we found that the first peak is not necessarily the highest peak in full profile due to nuisance strong reflectors in vicinity. 

\noindent \textbf{Object Classifier.}
Taking the extracted SOI as input, a classifier is used to identify a target object. The classifier adopted by \sysname is a convolution neural network (CNN), which is widely used in many classification tasks for its superior accuracy and efficiency. Specifically, this classifier comprises of three 1D convolution layers and a dense layer with softmax activation. The kernel sizes and strides of all three convolution layers $32$ and $1$, and the activation functions are Exponential Linear Unit (ELU). We compare the performance of this CNN classifier with other baseline classifiers in Sec.~\ref{sub:grid_map_reconstruction_performance} to further justify our choice.

%!TEX root = ../main.tex

\section{Implementation} % (fold)
\label{sec:implementation}

For the purpose of reproducing our approach, we release our dataset and the source code at \url{https://github.com/ChristopherLu/milliMap}.

\noindent \textbf{Multi-modal Robotic Sensing Platform.}
A Turtlebot 2 platform equipped with multiple sensors is used a a prototype data collection platform.
This dataset contains synchronized mmWave point cloud data from a TI AWR1443 board, lidar data from a Velodyne VLP-16 and wheel odometry. The bandwidth of the used radar is $4$GHz (77GHz - 81GHz) which yields a ranging resolution of $\sim 4$cm. It has $120$ degree azimuth field of view and $30$ degree elevation field of view. 
In addition, we provide RGB images from a front-facing monocular camera. The mmWave sensor, lidar and camera are coaxially located on the robot along the vertical axis.
The navigation of the mobile robot is implemented using ROS \cite{quigley2009ros} on a Linux notebook, which is a widely adopted practice in the robotics community. Besides controlling, the notebook is also responsible for sensor data storage. Once the collection phase is completed, the notebook sends the collection back to a backend server for offline model training. During the online phase, model inference is expected to be done either by an embedded GPU or the notebook itself. We will discuss the real-time performance soon in Sec.~\ref{sub:system_efficiency}.

\noindent \textbf{Testbeds.}
Two buildings are surveyed at the time of writing. The \textit{\buildA} Building has a size of $\sim 1,100m^2$ and contains four floors, mostly composed of corridors and atrium; the \textit{\buildB} Building has a size of $\sim205m^2$ and contains one floor with a combination of corridors and rooms. The \textit{\buildA} Building dataset presents a combination of walls, doors, lifts and large glass handrails; the \textit{\buildB} Building dataset presents walls, doors, glass panes, lifts and clutter. Notably, despite similar high-level semantics, these buildings differ in pathway widths, door types, glass sizes and more importantly, layouts. 

\noindent \textbf{Data Collection Procedure.} To collect the dataset of map reconstruction, we use a remote control to drive our mobile robot moving from a starting point to an end point on each floor of the buildings. Particularly, we do not set any specific traveling routes in data collection, but let the robot freely traverse the indoor space. The reconstruction dataset contains the data from the mmWave radar, lidar and wheel odometry. Sec.~\ref{sub:grid_map_reconstruction_performance} introduces how the collected data are used for training and testing. The semantic mapping dataset is acquired in the same places as above. In data collection, a mmWave radar on the robot is firstly rotated to a pose perpendicular to the target object/material surface with a distance $\sim 0.5$ meter.  Then at each collection point, we acquire data at a rate of $10$Hz and semantically label these offline from location logs. In total, we collected $45,535$ frames from $4$ types of objects in two buildings.

%!TEX root = ../main.tex
\section{Experimental evaluation}
\label{sec:experiments}

\subsection{Grid Map Reconstruction Performance} % (fold)
\label{sub:grid_map_reconstruction_performance}

We start with the validation of the grid map reconstruction method proposed in Sec.~\ref{sec:grid_map_reconstruction}.

\noindent \textbf{Evaluation Metrics.}
Throughout this section, two metrics are consistently adopted to quantify map reconstruction performances: mean absolute error ($L_1$) and mean \emph{intersection-over-union (IoU)}, both of which are widely used \cite{weston2018probably}. The mean $L_1$ is calculated as follows \cite{zhao2016loss}:
\begin{equation}
	L_1 = \frac{1}{N} \sum_{p \in P} |x(p)-y(p)|
\end{equation}
where $p$ is the index of the pixel and $P$ is the patch. $x(p)$ and $y(p)$ are the values of the pixels in the processed patch and the ground truth respectively.
We will omit ``mean'' hereafter for presentation ease. It is worth mentioning that as the image resolution is $1$dm/pixel in our case, \emph{the $L_1$ mapping error is thus in the units of decimeters}. 
It is also worth mentioning that our goal is to build an indoor map for navigation in search and rescue applications. Therefore it is necessary to have a good idea of the free space and obstacles. Although this property is difficult to be numerically reported, we will qualitatively discuss it when comparing reconstruction results.

\noindent \textbf{Evaluation Protocol.}
We perform cross-floor and cross-building tests to examine the effectiveness of the trained model. To avoid the known overfitting issues of DNN in our model and we particularly follow this cross-test evaluation principle on unseen scenarios.
Concretely, our collected dataset is divided into training and testing sets. In particular, the training set contains $12,000$ augmented patch images extracted from maps of the 1st, 2nd and 3rd floors in \buildA Building. The data augmentation strategy we adopt here is the standard rotation and translation transformations on original patches to promote model generalization. Our test set comprises $49$ patch images extracted from maps of the 4th floor in \buildA Building and $12$ patches extracted from the 2nd floor of \buildB Building.As introduced in Sec.~\ref{sec:implementation}, the environments of \buildA Building and \buildB Building notably differ in pathway widths, door types, glass sizes and more importantly, layouts etc.Moreover, the path followed by our robot on the 4th floor is quite different from that of other three floors in \buildA Building. The above scenario variety helps us maximally follow the cross-testing principle.

All training and testing patch images have size $64 \times 64$. Concerning model training, three loss weights $\lambda_1$, $\lambda_2$ and $\lambda_3$ are set to $10$, $10$ and $5$ respectively. We adopt a line detector as the convolution kernel in Eq.~(\ref{eq: mp_loss}), $M$ is set to $4$, corresponding to $4$ line directions in \ang{0}, \ang{45}, \ang{90} and \ang{135}. The training batch size is set to $16$ and we use the Adam optimizer at a learning rate of $2e^{-3}$.

\begin{table}[!t]
\caption{Densification Before and After Mapping.}
\label{tab:scan_vs_patch}
	\centering
	\small
	\begin{tabular}{|c|c|c|c|c|c|}
	\hline
	\multirow{2}{*}{\textit{}} & \multirow{2}{*}{\textit{Method}} & \multicolumn{2}{c|}{\textit{\buildA Building}} & \multicolumn{2}{c|}{\textit{\buildB Building}} \\ \cline{3-6} 
	 &  & L1 & IoU & L1 & IoU \\ \hline
	\multirow{2}{*}{\begin{tabular}[c]{@{}c@{}}Scan\\ (before)\end{tabular}} & Pix2Pix \cite{isola2017image} & 2.776 & 0.186 & 3.602 & 0.150 \\ \cline{2-6} 
	 & Pix2PixHD \cite{wang2018high} & 2.309 & 0.226 & 3.722 & 0.152 \\ \hline
	\multirow{2}{*}{\begin{tabular}[c]{@{}c@{}}Patch\\ (after)\end{tabular}} & Pix2Pix \cite{isola2017image} & 2.214 & 0.319 & 3.200 & 0.173 \\ \cline{2-6} 
	 & Pix2PixHD \cite{wang2018high} & \textbf{2.096} & \textbf{0.380} & \textbf{2.752} & \textbf{0.239} \\ \hline
	\end{tabular}
\vspace{-0.4cm}
\end{table}

\noindent \textbf{Effectiveness of Densification Before and After Mapping.}
We first investigate the effect of two input representations (refer to Section \ref{sub:network_input}):
(i) we perform densification of each scan and then aggregate them using grid mapping (denoted as \emph{scan} representation) and (ii) we aggregate scans using grid mapping and then perform densification on image patches (denoted as \emph{patch} representation). As Tab.~\ref{tab:scan_vs_patch} shows, the reconstruction results of \emph{patch} representation are significantly better than \emph{scan} for both networks, implying the effectiveness of \emph{patch} representation. Given the best-performing Pix2PixHD network, the $L_1$ errors of \emph{scan} are $20\%$ inferior to \emph{patch}, with over $35\%$ inferior IoU scores on both datasets. The reason is that the single \emph{scan} densification easily overfits to straight lines, which is consistent with our discussion in Sec.~\ref{sub:network_input}. 

% the order impacts of scan densification and stitching. With two reconstruction neural networks, we compare their performance under different processing orders: (i) scan densification first and then stitching (denoted as \emph{scan} representation) and (ii) scan stitching first and then densification (denoted as \emph{patch} representation). As we can see in Tab.~\ref{tab:scan_vs_patch}, the reconstruction results of \emph{patch} is significant better than \emph{scan} for both networks, implying the effectiveness of the \emph{patch} method. Given the best-performing Pix2PixHD network, the $L_1$ errors of \emph{scan} are $20\%$ inferior to \emph{patch}, with a half IoU scores achieved on both datasets. The network outputs of \emph{scan} are easily overfit to straight lines, which is consistent to our discussion in Sec.~\ref{sub:network_input_representation}. 

\begin{figure*}[t]
\centering
  \includegraphics[width=\linewidth]{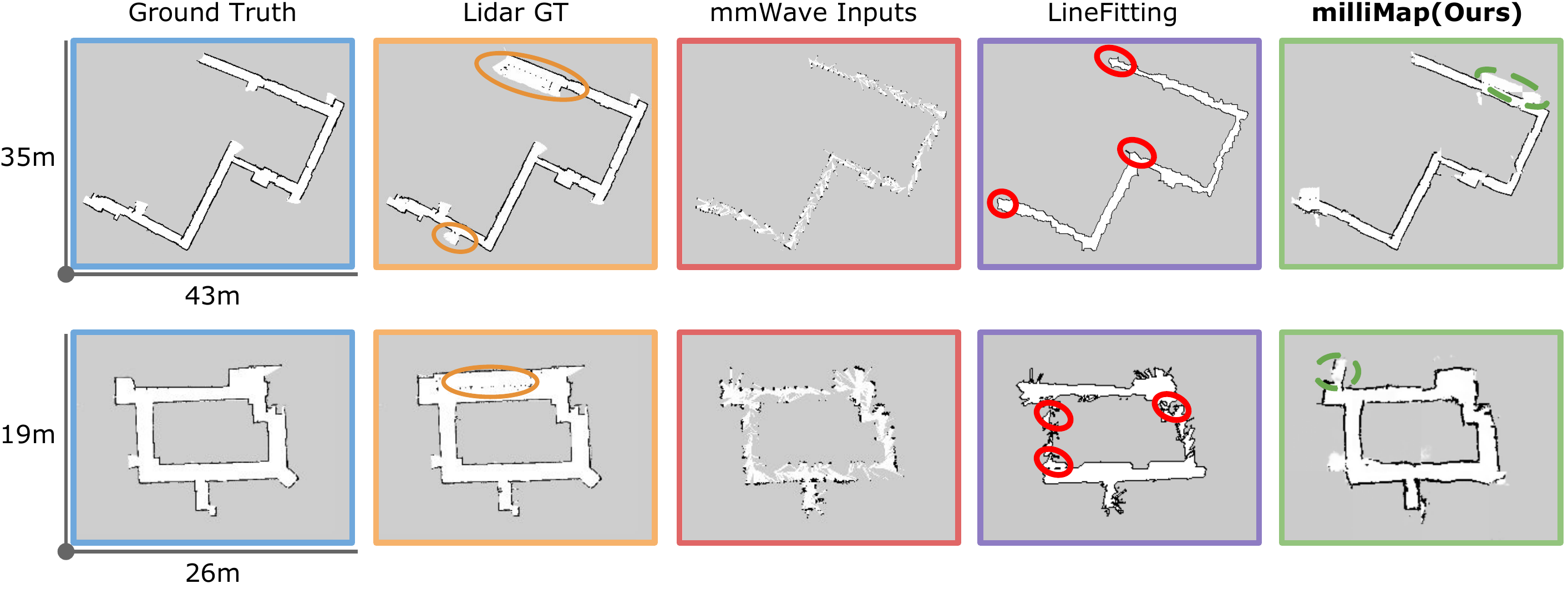}
  \caption{Qualitative reconstruction results. \sysname achieves a comparable performance to the lidar counterpart. Solid circles on Lidar GT are glass objects; dashed circles are `ghost areas' in generation. Red circles show corridors that have been erroneously closed by the line-fitting method (\emph{false obstacles}). Top Row: \buildA Building; Bottom Row: \buildB Building.}
  \label{fig:map_reconstruct_compare}
\end{figure*}

\begin{figure}[!t]
\centering
  \includegraphics[width=\linewidth]{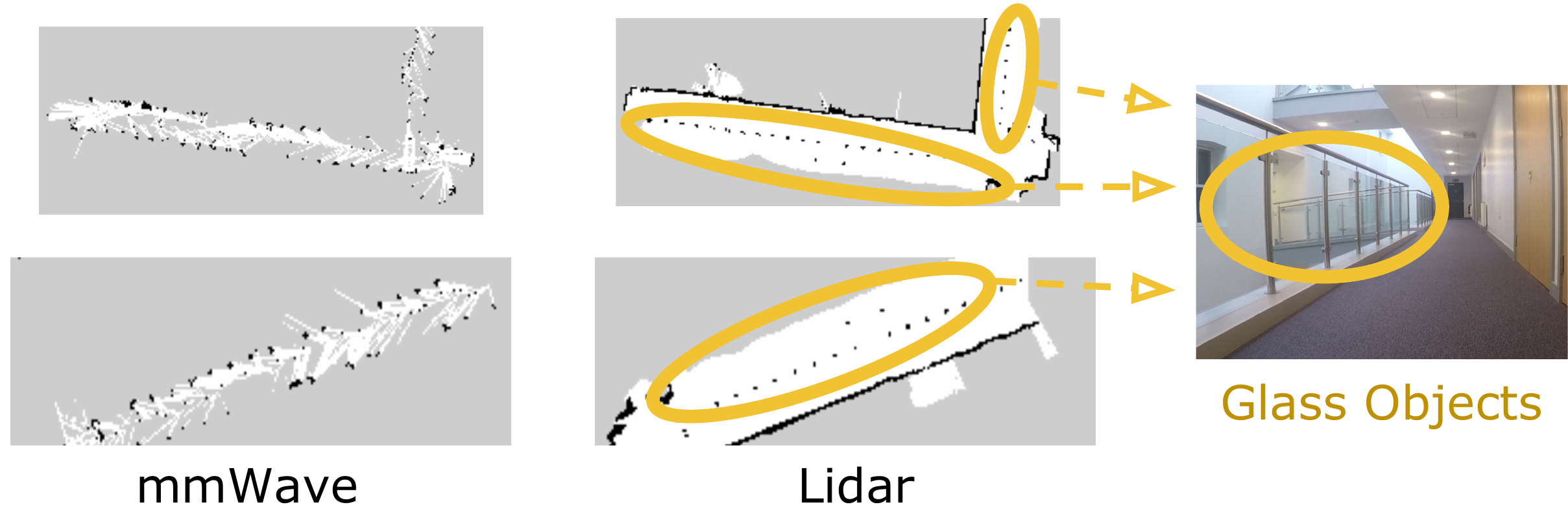}  %0.97
  \caption{Incorrect lidar supervision due to presence of glass objects in training data.}
  \label{fig:wrong_supervision}
  	\vspace{-0.4cm}
\end{figure}

\begin{figure*}[!t]
\centering
  \includegraphics[width=0.95\linewidth]{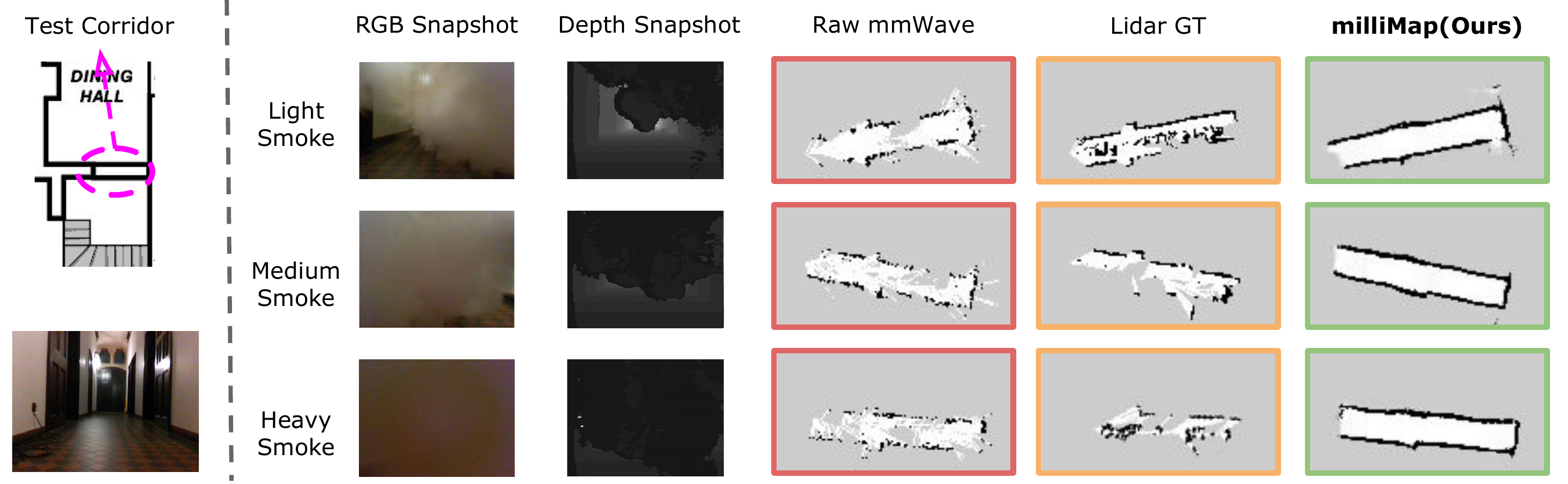}  %0.97
  \caption{Qualitative testing in smoke-filled environments.}
  \label{fig:smoke_map}
  	%\vspace{-0.5cm}
\end{figure*}

\noindent \textbf{Network Architecture Validation}
After understanding the effective processing order, we adopt the \emph{patch} representation for subsequent	experiments and continue to validate different architectures of reconstruction networks. As \sysname is the first indoor mapping work dealing with very sparse inputs of such low-cost mmWave radar, we can only compare the following commonly used generative networks: Conditional Variational Autoencoder (CVAE) \cite{weston2018probably}, BicycleGAN \cite{zhu2017toward}, Pix2Pix \cite{isola2017image} and Pix2PixHD \cite{wang2018high}.
Notably, CVAE is the network architecture adopted by \cite{weston2018probably}, though their goal is not sparse-to-dense due to the use of a customized mechanical radar. 
Beside these deep learning methods, we also compare with lineFitting \cite{pfister2003weighted}, a classic reconstruction method for line-based indoor floor plans.
Tab.~\ref{tab:map_reconstruction} shows the performance comparison of different reconstruction methods. Despite its success on lidar map reconstruction, the classic line fitting method obviously struggles on both datasets and provides $<50\%$ IoU than our approach, attributed to the substantial sparsity in raw mmWave maps. In particular, it is observed in Fig.~\ref{fig:map_reconstruct_compare} that there are many falsely closed corridors predicted by the line fitting method. Such misclassified free space and navigable routes is contrary to our goal for safe/efficient navigation as areas falsely marked as obstacles are in general more detrimental than areas falsely marked as free space, since a robot or a firefighter is typically capable of avoiding unpredicted obstacles. In contrast, when computing a path to a certain location, falsely closed corridors could make whole areas of the building appear inaccessible.
On the side of DNN methods, we did not find the advantages of using variational methods, implying that random sampling from a learnt distribution actually counteracts the benefits of uncertainty modelling and tends to output blurred reconstructions. We hypothesize that the performance gain can be also attributed to the strong regularity within indoor maps, which favors deterministic learning methods. Lastly, despite their close correlation, we found that Pix2PixHD outperforms Pix2Pix on both datasets, thanks to the use of multi-scale discriminators and more losses. By introducing the map-prior loss, our method can further gain $9.6\%$ L1 accuracy than Pix2PixHD, and $\sim 5\%$ better IoU performance overall on both datasets, which is a comparable delta to the field of image reconstruction/translation \cite{zhu2017unpaired}. Note that the prior loss is simply an additional loss term that incurs no further computation overhead for either inference or training; however, it still leads to a performance increase.

\noindent \textbf{Explanation of `Ghost' Areas.}
Interestingly, in the last column of Fig.~\ref{fig:map_reconstruct_compare}, there are `ghost' areas on the generated maps, where part of a wall (black) is incorrectly marked as free regions (white). Recall that we adopt a cross-modal supervised learning framework that uses lidar patches as supervision labels. These labels, however, can be error-prone when encountering glass objects (see the second column in Fig.~\ref{fig:map_reconstruct_compare}), which is a commonly-known limitation of lidar. Although glass is opaque to mmWave, considering the high appearance similarity (see Fig.~\ref{fig:wrong_supervision}), we hypothesize the `ghost area' of our generated grid map of \buildA Building can be attributed to the misleading lidar patches of glass in training. `Ghost' areas do not appear with scan inputs, due to its overfitting to straight corridors.

\begin{table}[!t]
	\caption{Reconstruction method comparison.}
	\label{tab:map_reconstruction}
	\centering
	\small
	% \begin{tabular}{|c|c|c|c|c|c|}
	% \hline
	% \multirow{2}{*}{\textit{Input}} & \multirow{2}{*}{\textit{Model}} & \multicolumn{2}{c|}{\textit{Wolfson}} & \multicolumn{2}{c|}{\textit{RHB}} \\ \cline{3-6} 
	%  &  & L1 & IoU & L1 & IoU \\ \hline
	% Scan & Pix2PixHD \cite{wang2018high} & 2.316 & 0.215 & 2.731 & 0.136 \\ \hline
	% \multirow{4}{*}{Patch} & CVAE \cite{sohn2015learning}& 2.408 & 0.323 & 3.082 & 0.221 \\ \cline{2-6} 
	%  & BicycleGAN \cite{zhu2017toward} & 2.538 & 0.303 & 3.393 & 0.195 \\ \cline{2-6} 
	%  & Pix2Pix \cite{isola2017image} & 2.214 & 0.319 & 3.200 & 0.173 \\ \cline{2-6} 
	%  & Ours & \textbf{1.931} & \textbf{0.398} & \textbf{2.589} & \textbf{0.238} \\ \hline
	% \end{tabular}
	\begin{tabular}{|c|c|c|c|c|}
	\hline
	\multirow{2}{*}{\textit{Method}} & \multicolumn{2}{c|}{\textit{\buildA Building}} & \multicolumn{2}{c|}{\textit{\buildB Building}} \\ \cline{2-5} 
	 & L1 & IoU & L1 & IoU \\ \hline
	LineFitting \cite{pfister2003weighted}  & 3.180 & 0.167 & 4.114 & 0.103 \\ \hline
	CVAE \cite{weston2018probably} & 2.408 & 0.323 & 3.082 & 0.221 \\ \hline
	BicycleGAN \cite{zhu2017toward} & 2.538 & 0.303 & 3.393 & 0.195 \\ \hline
	Pix2Pix \cite{isola2017image} & 2.214 & 0.319 & 3.200 & 0.173 \\ \hline
	Pix2PixHD \cite{wang2018high} & 2.096 & 0.380 & 2.752 & 0.239 \\ \hline
	Ours & \textbf{1.976} & \textbf{0.402} & \textbf{2.536} & \textbf{0.247} \\ \hline
	\end{tabular}
	\vspace{-0.4cm}
\end{table}% (fold)

\subsection{Effectiveness of Sub-components} % (fold)
\label{sub:effectiveness_analysis_of_components}

In order to understand the contribution of key sub-components in the reconstruction neural network, we further conduct an effectiveness analysis on: i) loss functions and ii) multi-scale discriminators. 

\noindent \textbf{Different Loss Functions.}
We modify the objective function of Eq.~\ref{eq: total_loss}, by alternating different loss terms for reconstruction likelihood as well as alternating variants of our proposed map-prior term. Tab.~\ref{tab:ablation} shows that feature matching loss plays a vital role which brings $16\%-24\%$ gain in $L_1$. 
The perceptual loss (i.e., VGG loss) also helps and removing it incurs a average performance decline ($\sim 7\%$) on both datasets. This is reasonable because the VGG network is pre-trained by general image classification tasks and hence becomes less effective in our specific mapping task. 

These experiments indicate that, although grid maps are more about geometrics, these appearance losses are still important for stabilising generator training and improving realism.
Interestingly, when we implement the map prior loss as edge detectors, its efficacy is not as helpful as the line detectors. This is because edges are a broad concept for any image and cannot effectively incorporate the geometrics of line-based maps. Moreover, as our supervision signals are from the imperfect lidar patches, the edge detectors are sensitive to the noises of lidar. In contrast, line detectors focus on low-frequency components of images and thus can be more robust to noise.

\noindent \textbf{Number of Scales.}
Next we examine the impact of multi-scale discriminators. Recall that \sysname uses a $2$-scale discriminator while our ablation study further examines the cases of $1$- and $3$-scales. As shown in Tab.~\ref{tab:ablation}, the overall impact of multi-scale discriminators is not substantial ($\sim 5\%$) when varying the number of scales. This is as expected because the multi-scale discriminators were originally designed for high-resolution images while our input patches are not. We observed a marginal improvement from single-scale to $2$-scale discriminators as more diverse feature matching is introduced in different scales. However, such increase of scales soon counteracts the benefits when the $3$-scale network becomes oversized and overfits. This overfitting issue is more obvious on \textit{\buildB Building} dataset due to cross-building testing.

\subsection{Testing in Smoke-filled Environments} % (fold)
\label{sub:testing_in_smoke_filled_environments}

Thick smoke is a common event that occurs in many emergency incidents such as firefighting. 
In this experiment we examine the potential use of \sysname in smoke-filled environments. To this end, we use a smoke machine to create different smoke densities in a corridor ($12 \times 1.5$m$^2$) in another building where various sensor data were collected on the robotic platform for comparison, including lidar, depth camera, RGB-camera and mmWave radar.
Fig.~\ref{fig:smoke_map} shows the reconstructed map in $3$ scenarios with different levels of smoke distributions. As we can see, lidar gives very inaccurate map results even with low levels of smoke. Due to the occlusion and reflection effects of smoke particles, lidar generates many non-existent objects and/or misses a lot of real ones. In fact, even under the lightest smoke condition, lidar already undergoes substantial performance degradation. Depth and RGB cameras also fails to see through smoke due to similar reasons.
In contrast, the mmWave radar is able to see through smoke and \sysname reconstructs the corridor accurately in all $3$ smoke-filled scenarios. These results demonstrate that our mmWave based reconstruction model trained in benign environments can transfer its mapping ability to unseen smoke-filled environments. Based on this trial, we believe there are many promising use cases of it for emergency situations.

\begin{table}[!t]
	\caption{Effectiveness on losses and number of scales.}
	\label{tab:ablation}
	\centering
	\small
	\begin{tabular}{|c|c|c|c|c|c|}
		\hline
		\multicolumn{2}{|c|}{\multirow{2}{*}{\textit{}}} & \multicolumn{2}{c|}{\textit{\buildA Building}} & \multicolumn{2}{c|}{\textit{\buildB Building}} \\ \cline{3-6} 
		\multicolumn{2}{|c|}{} & L1 & IoU & L1 & IoU \\ \hline
		\multirow{3}{*}{Losses} & w.o. FM & 2.408 & 0.323 & 3.082 & 0.221 \\ \cline{2-6} 
		 & w.o. VGG & 2.115 & 0.379 & 2.762 & 0.242 \\ \cline{2-6} 
		 & Edge Loss & 2.214 & 0.319 & 3.200 & 0.173 \\ \hline
		\multirow{2}{*}{\begin{tabular}[c]{@{}c@{}}\# of\\ Scales\end{tabular}} & 1 & 2.024 & 0.394 & 2.633 & \textbf{0.250} \\ \cline{2-6} 
		 & 3 & 2.022 & 0.387 & 2.863 & 0.219 \\ \hline
		\multicolumn{2}{|c|}{Ours} & \textbf{1.931} & \textbf{0.398} & \textbf{2.589} & 0.238 \\ \hline
	\end{tabular}
	\vspace{-0.4cm}
\end{table}

\begin{figure}[!t]
\centering
  \includegraphics[width=0.95\linewidth]{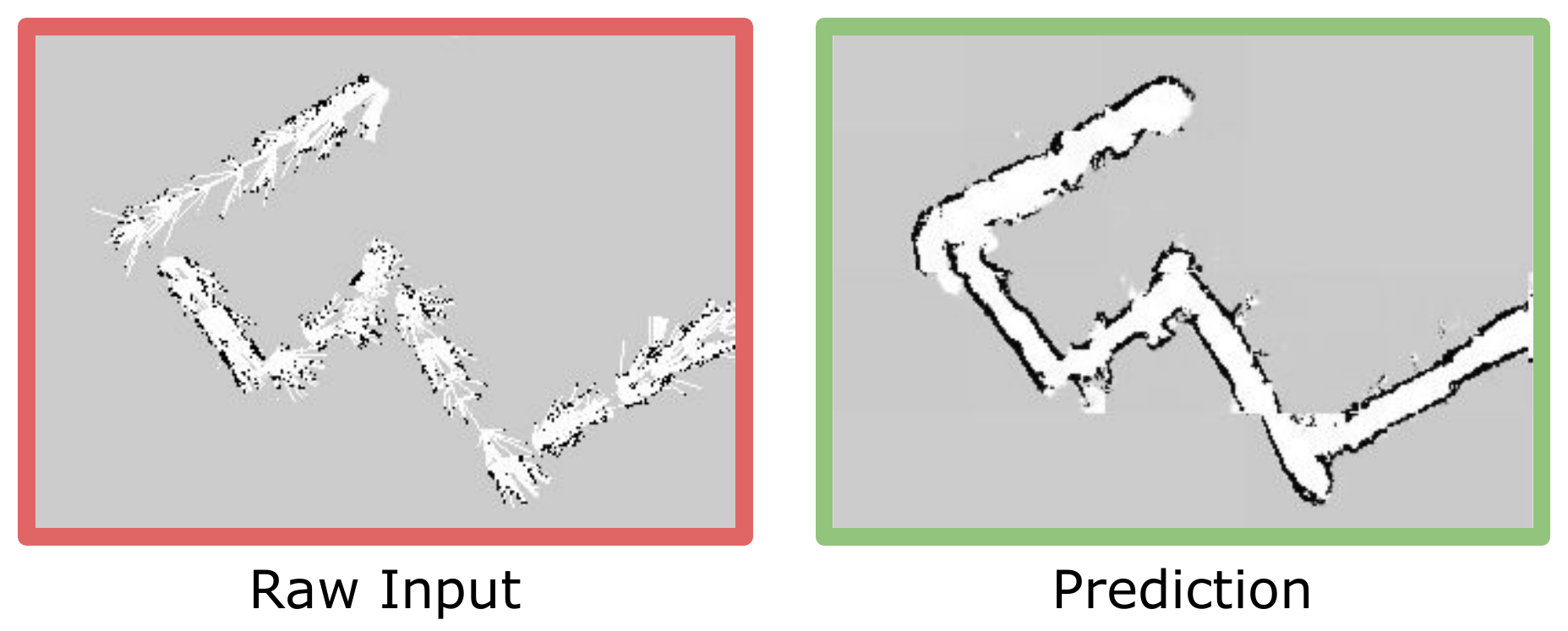}
  \caption{Qualitative result for hand-held cases.}
  \label{fig:map_pdr}
  	\vspace{-0.5cm}
\end{figure}

\subsection{Extending to Hand-held Devices} % (fold)
\label{sub:extending_to_hand_held_devices}

First responders, who carry hand-held or helmet-mounted devices, need to work in a team with robots for complementary operations. To this extent, we test \sysname's potential for map construction on hand-held devices, without retraining, but directly using the model trained using a robot. The main differences are that the odometry of the hand-held device is inferred from an embedded inertial measurement unit by pedestrian dead reckoning (PDR) methods \cite{jimenez2009comparison}. However, compared to wheel odometry, PDR odometry drifts more and has a lower sampling rate due to step discretization. As a consequence, the raw patch images of PDR are of lower fidelity. Furthermore, due to different viewpoints (e.g., different heights of robots and pedestrians), the mmWave observations have obvious differences from the training samples. Despite these compromising factors, as can be seen in Fig.~\ref{fig:map_pdr}, \sysname still gives a good reconstruction with $\sim 0.83m$ error, providing a much better sense of space accessibility than using raw data alone. This experiment serves to demonstrate how teams of robots and people could build a common map. 

\subsection{Downstream Navigation Tasks} % (fold)
\begin{figure}[!t]
	\centering
	\begin{subfigure}[b]{0.235\textwidth}\centering
		\includegraphics[width=\columnwidth]{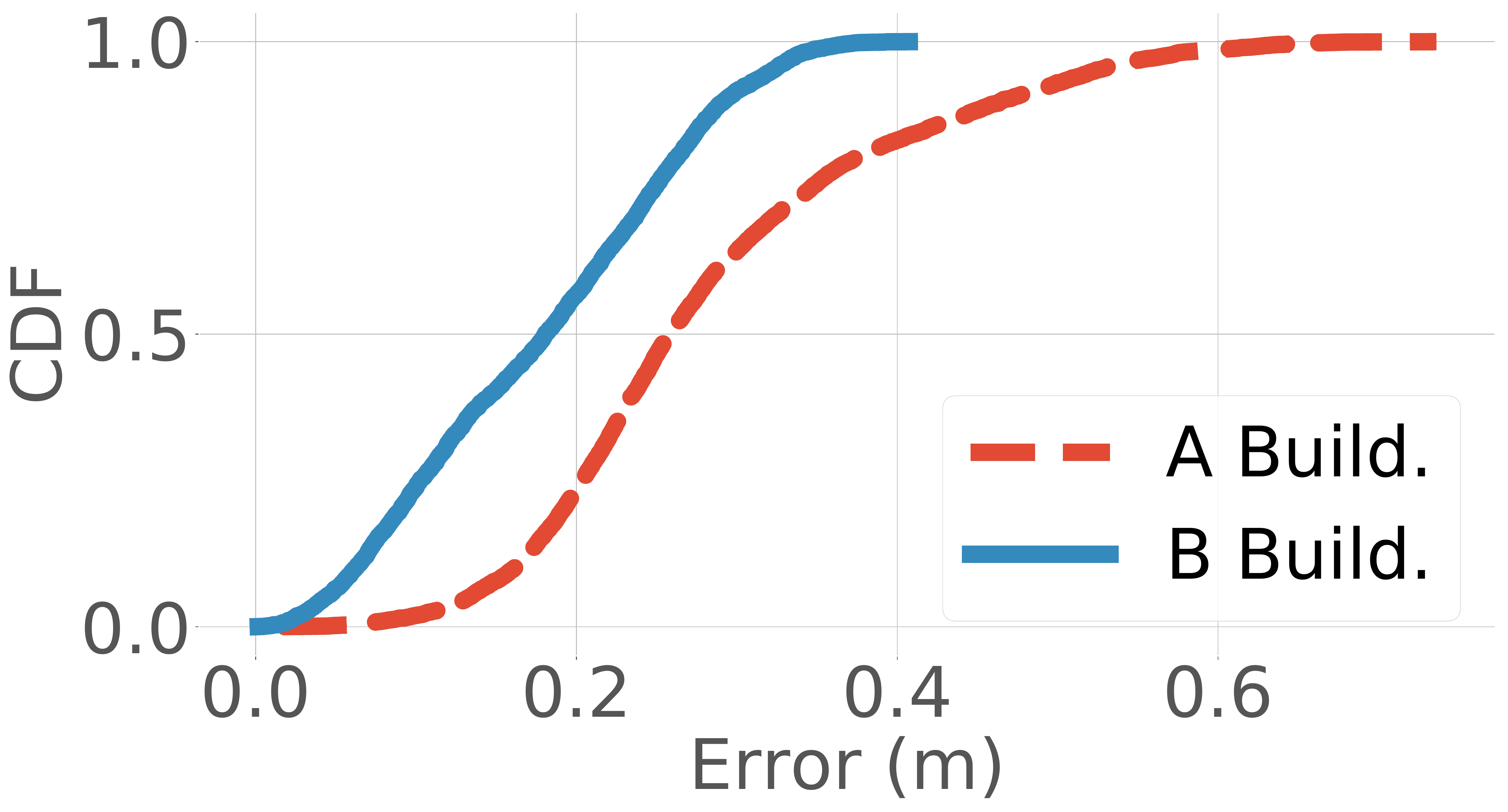} 
		\caption{Translation}
		\label{fig:trans_errorn}
	\end{subfigure}
	\hfill
	\begin{subfigure}[b]{0.235\textwidth}\centering
		\includegraphics[width=\columnwidth]{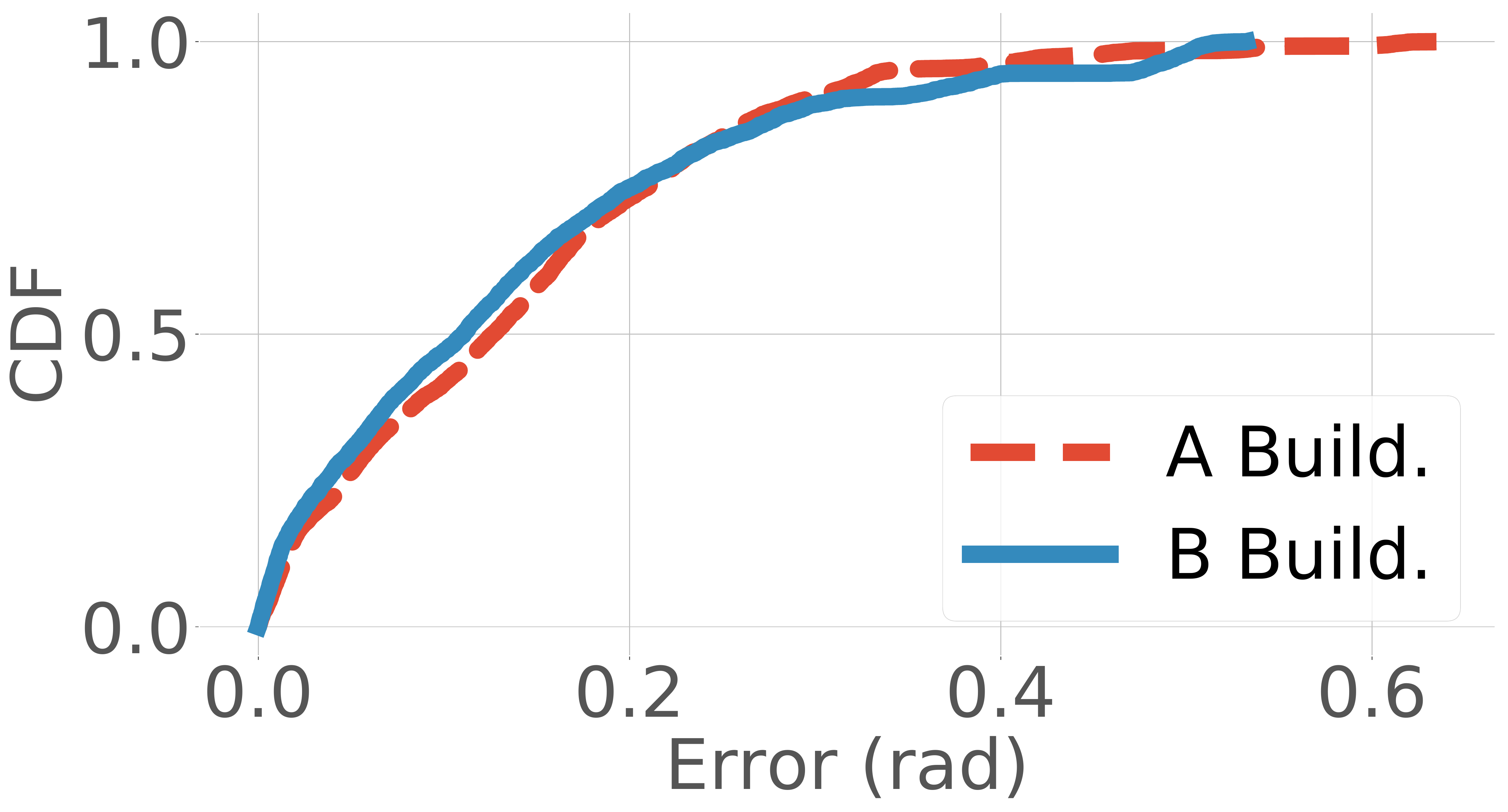} 
		\caption{Orientation}
		\label{fig:ori_error}
	\end{subfigure}%
\caption{Error CDFs for the downstream localization tasks.}
\label{fig:err_cdf}
\end{figure}

\label{sub:downstream_task_performances}
We now test whether the produced maps, despite their imperfections, can still be used for autonomous navigation. In particular, we investigate if another robot or person is able to localize in the predicted map with comparable accuracy to that of a lidar map. We run Monte Carlo localization using mmWave raw measurements on the reconstructed maps using the standard \textit{amcl} ROS package with default parameters. Each time the robot or person starts at a random location and samples a radar frame. The pseudo-ground truth is derived by localization with lidar on a lidar map of the same floor. Fig.~\ref{fig:err_cdf} shows the cumulative error distribution for $50$ Monte Carlo runs. For the reconstructed map of \buildA Building, our robot achieved a mean translation accuracy of $0.285$m and orientation accuracy of $0.142$ rad; on the reconstructed map of \buildB Building, the mean translation and orientation accuracy are $0.178$m and $0.140$ rad respectively. Given the size of the two buildings, these results show that the map produced by \sysname can be used to accurately localize and navigate firefighters or robots.

\subsection{Semantic Mapping Performance} % (fold)
\label{sub:semantic_mapping_performance}

\noindent \textbf{Metrics and Baselines.} 
To validate the performance of semantic classification, we adopt the $4$ metrics for standard classification tasks: \emph{accuracy, precision, recall and F1 score}. For comparison, we implement RSA \cite{zhu2015reusing}, a method identifying objects based on the mmWave reflectivity on different surface materials. Furthermore, to justify our choice of CNN classifier, we also compare with other $4$ commonly used classifiers, including support vector machine (SVM), random forest (RF), k-nearest neighbors (KNN), multi-layer perceptron (MLP). All of these classifiers take as inputs SOI and predict an object label out of glass, lift, wall and door. 

\noindent \textbf{Evaluation Protocol.} % (fold)
The evaluation protocol here is similar to the one described in Sec.~\ref{sub:grid_map_reconstruction_performance}. Specifically, classifiers are developed on a training set collected from three floors in \buildA Building and we test the trained classifier on a new floor in \buildA Building as well as in a new building of \buildB. Overall, our training and test sets contain $27,952$ and $17, 583$ samples (two test buildings) respectively. When training baselines and our classifier, the best model for online inference is determined by 5-fold cross validation.

\begin{figure}[h]
	\centering
	\begin{subfigure}[b]{0.24\textwidth}\centering
		\includegraphics[width=\columnwidth]{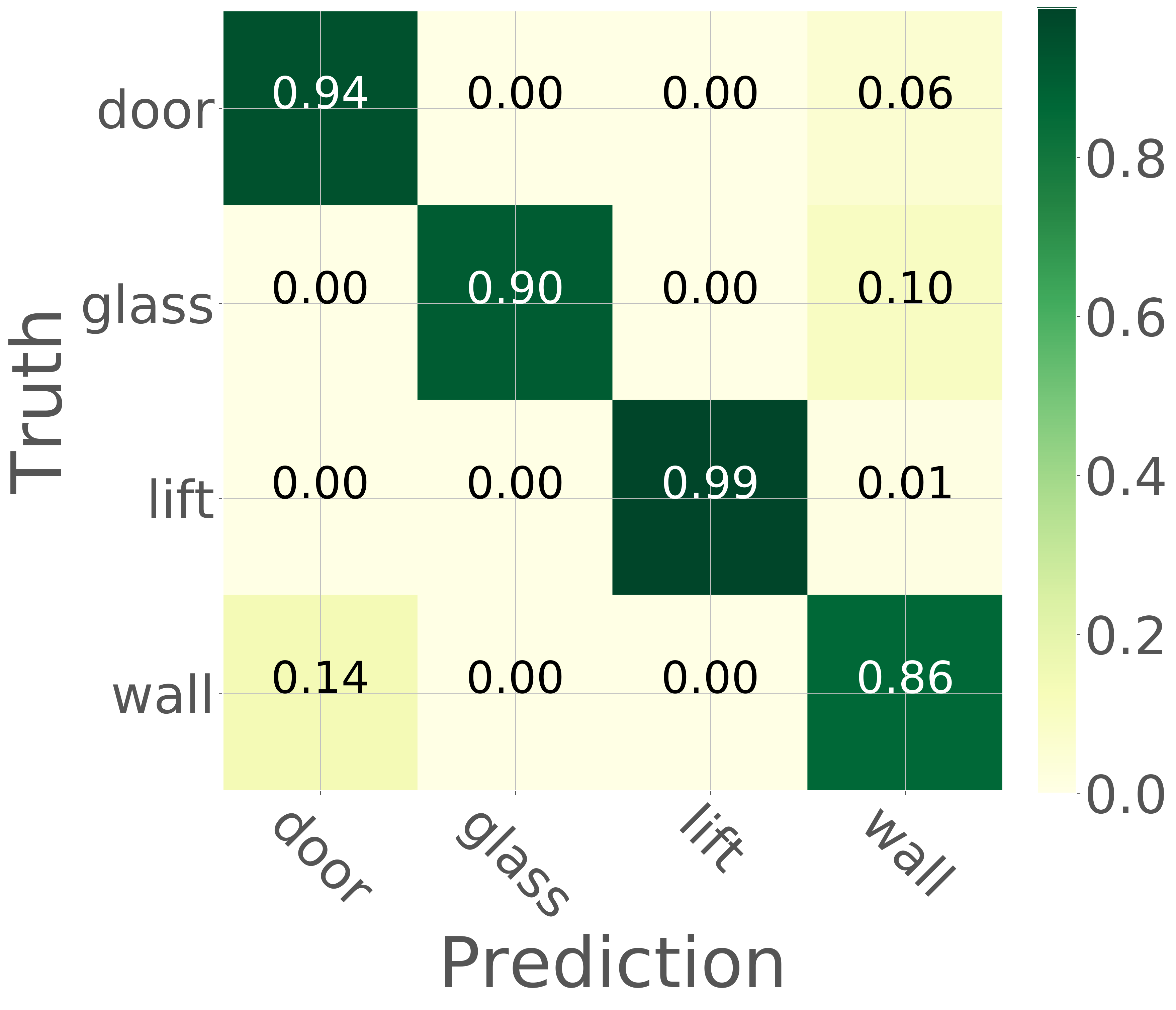} 
		\caption{\buildA Building}
		\label{fig:cnf_wolfson}
	\end{subfigure}%
	\begin{subfigure}[b]{0.24\textwidth}\centering
		\includegraphics[width=\columnwidth]{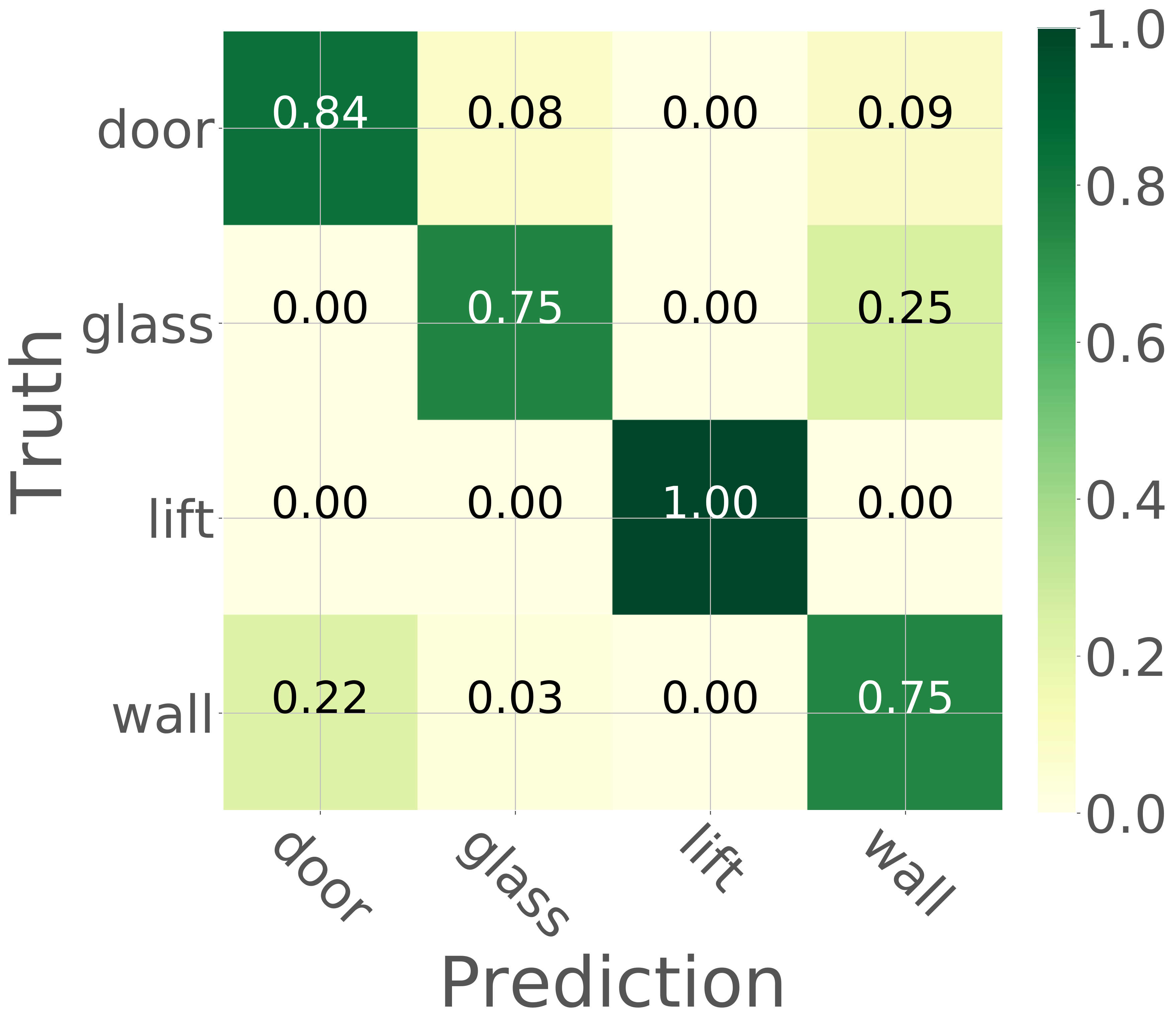} 
		\caption{\buildB Building}
		\label{fig:cnf_rhb}
	\end{subfigure}%
\caption{Confusion Matrix of CNN classifier: (a) \buildA Building (b) \buildB Building.}
\label{fig:cnf}
\end{figure}

\noindent \textbf{Overall Performance.} % (fold)
Tab.~\ref{tab:material_classification} summarizes the semantic mapping performance where a SOI with a width of $6$ is used. Clearly, our CNN classifier  achieves the best performance overall on two datasets and MLP classifier is second to it. All shallow-learning based classifiers (i.e., SVM, RF, KNN) underperformed relative to the deep-learning based methods. This is reasonable as MLP and CNN are able to learn meaningful feature representation in training, rather than a shallow classifier on raw data. Because of these meaningful features, MLP and CNN based classifiers can generalize across floors and buildings. In contrast, as RSA only considers the specular reflection from the surface material while ignoring the rich information conveyed by multi-path reflections, it struggles to robustly identify objects in both cases.
As expected, cross-building classification (\buildB Building Dataset) is more challenging than cross-floor classification (\buildA Building Dataset) because building differences are more substantial than floor differences, resulting in a performance gap $\sim 15\%$ on average. 
Fig.~\ref{fig:cnf} further plots the confusion matrix of our CNN classifier. We observed that walls are the most difficult objects to identify on both datasets, coinciding with its greater structural complexity than other objects. In contrast, lifts are generally made of steel, allowing them to be easily identified and yields very high accuracy.

\begin{table}[!t]
	\centering
	\small
	\caption{Results of Material Classification: Accuracy (Acc.), Precision (Prec.), Recall (Rec.) and F1 Score.}
	\label{tab:material_classification}
	\begin{tabular}{|c|c|c|c|c|c|c|c|c|}
	\hline
	\multirow{2}{*}{} & \multicolumn{4}{c|}{\textit{\buildA Building}} & \multicolumn{4}{c|}{\textit{\buildB Building}} \\ \cline{2-9} 
	 & Acc. & Prec. & Rec. & F1 & Acc. & Prec. & Rec. & F1 \\ \hline
	RSA & 0.67 & 0.74 & 0.69 & 0.71 & 0.50 & 0.58 & 0.53 & 0.56 \\ \hline
	KNN & 0.83 & 0.87 & 0.86 & 0.87 & 0.67 & 0.68 & 0.75 & 0.71 \\ \hline
	SVM & 0.82 & 0.86 & 0.85 & 0.85 & 0.67 & 0.70 & 0.68 & 0.69 \\ \hline
	RF & 0.86 & 0.89 & 0.89 & 0.89 & 0.67 & 0.68 & 0.72 & 0.70 \\ \hline
	MLP & 0.90 & 0.92 & \textbf{0.91} & 0.91 & 0.74 & 0.77 & 0.78 & 0.77 \\ \hline
	\emph{Ours} & \textbf{0.92} & \textbf{0.93} & 0.89 & \textbf{0.91} & \textbf{0.80} & \textbf{0.84} & \textbf{0.92} & \textbf{0.88} \\ \hline
	\end{tabular}
\end{table}

\begin{figure}[!t]
	\centering
	\begin{subfigure}[b]{0.24\textwidth}\centering
		\includegraphics[width=\columnwidth]{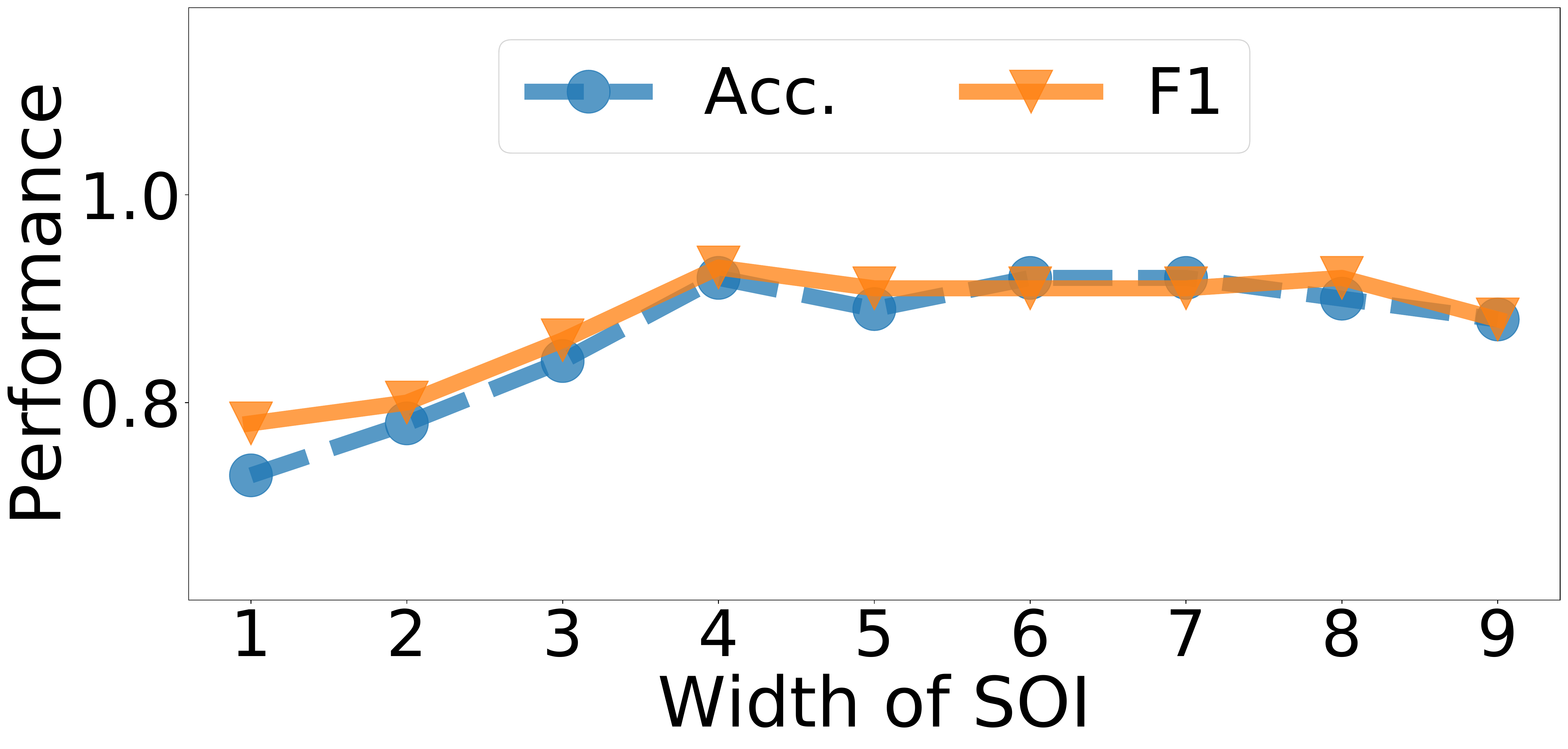} 
		\caption{\buildA Building}
		\label{fig:wolfson_soi_length}
	\end{subfigure}%
	\begin{subfigure}[b]{0.24\textwidth}\centering
		\includegraphics[width=\columnwidth]{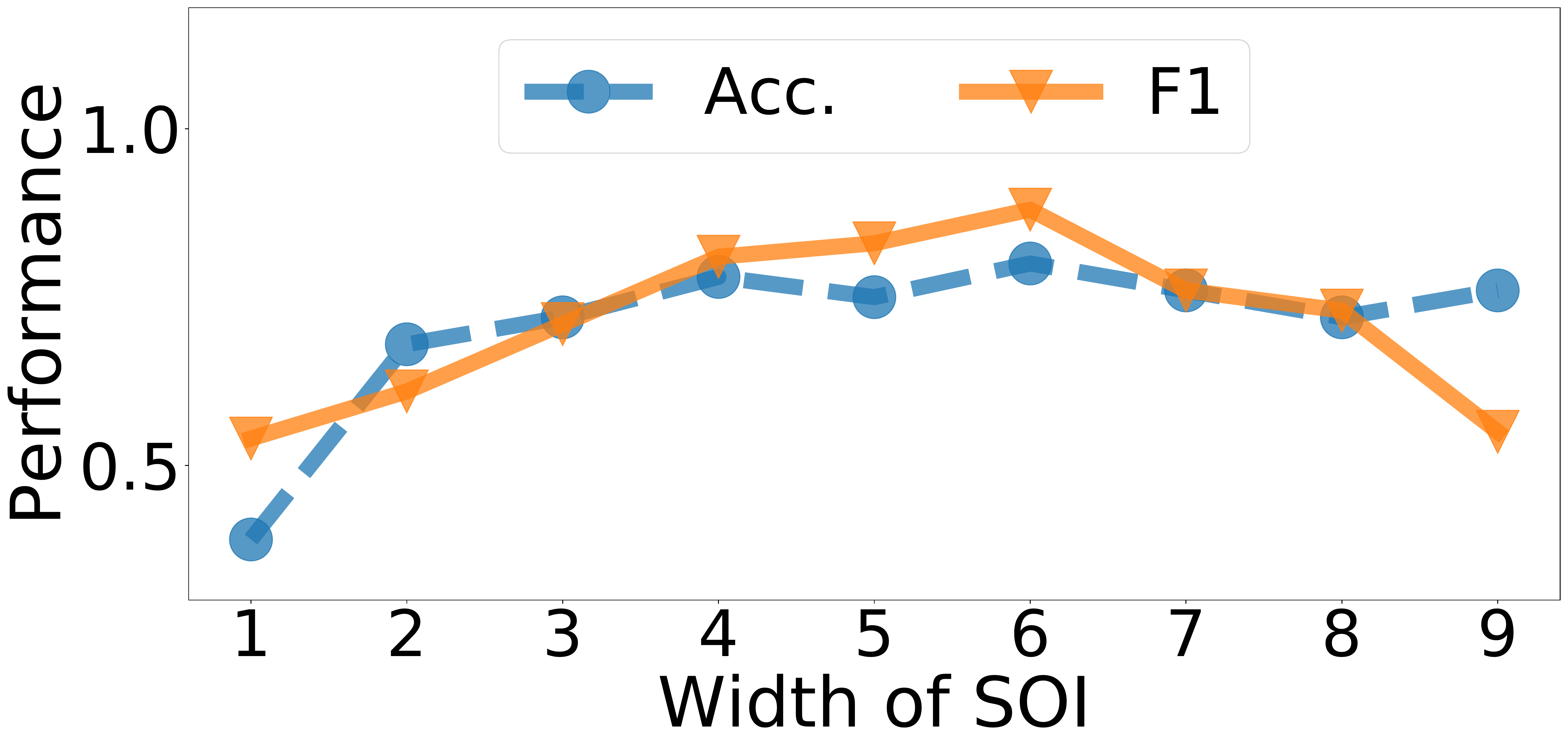} 
		\caption{\buildB Building}
		\label{fig:rhb_soi_length}
	\end{subfigure}%
\caption{Impact of the SOI width on semantic mapping.}
\label{fig:soi_length}
\vspace{-0.4cm}
\end{figure}

\noindent \textbf{Impact of SOI Length.} % (fold)
The width of SOIs is an important parameter which essentially determines the tradeoff between information richness of features and noise levels. To investigate its impact on the end-to-end object classification, we vary the width from $1$ to $9$, at a step of $1$. As we can see in Fig.~\ref{fig:soi_length}, an effective width falls into the range of $[4, 6]$ while either an over-long or over-short SOI results in a sub-optimal classification result. Notably, the negative impact of over-long SOIs is not as significant as the over-short case for unseen floors (see Fig.~\ref{fig:wolfson_soi_length}). We hypothesize that this is attributed to the adopted CNN which likely learns to suppress extraneous information of non-target reflections and such extraneous noise is similar across floors in the same building. However, the limitation of over-long SOIs becomes significant in the case of an unseen building, as suggested by the drop of F1 score in Fig.~\ref{fig:rhb_soi_length}. This is reasonable because more different secondary reflections are experienced due to the distinct building structures which makes the learned suppression hard to generalize. Empirically, SOIs with the width of $6$ gives the best overall performance.

\noindent \textbf{Dealing with Out-of-set Objects.} % (fold)
In real-world applications, it is possible that some objects or materials are not included in the training database, known as out-of-set or foreign/alien objects, and could cause false detections. To detect and mitigate their impacts on our semantic mapping, we introduce an `unknown' label to mark these out-of-set classes. Inspired by the `alien device' detection technique in \cite{li2018printracker}, we take the maximum probability value from the class distributions of softmax output (see Sec.~\ref{sub:semantic_recognizer}) as a classification score. To distinguish an unknown object from the known ones, we apply a threshold on the classification score - if the score is less than the threshold, we mark the object as unknown. The rationale behind such a score threshold is based on the principles of network learning and that the summation of a softmax distribution is always equal to $1$. Indeed, the goal of learning a CNN classifier is to maximize the softmax probability for individual true classes while a flat probability distribution over multiple classes in testing time often implies an out-of-set label.  

\begin{figure}[!t]
	\centering
	\begin{subfigure}[b]{0.47\textwidth}\centering
		\includegraphics[width=\columnwidth]{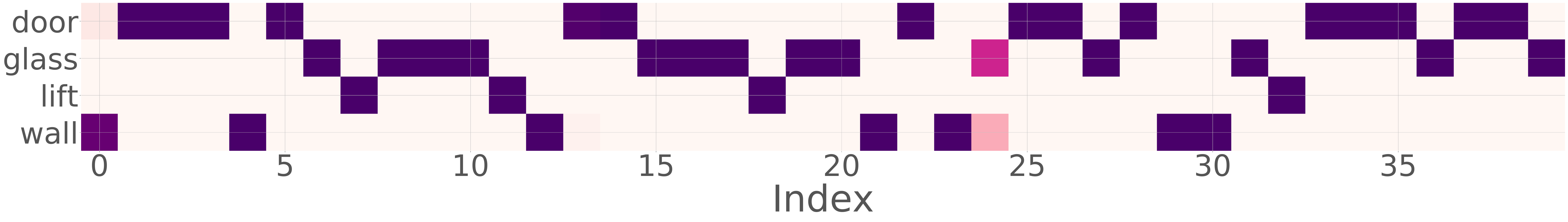} 
		\caption{Various known (in-database) classes.}
	\end{subfigure}%

	\begin{subfigure}[b]{0.47\textwidth}\centering
		\includegraphics[width=\columnwidth]{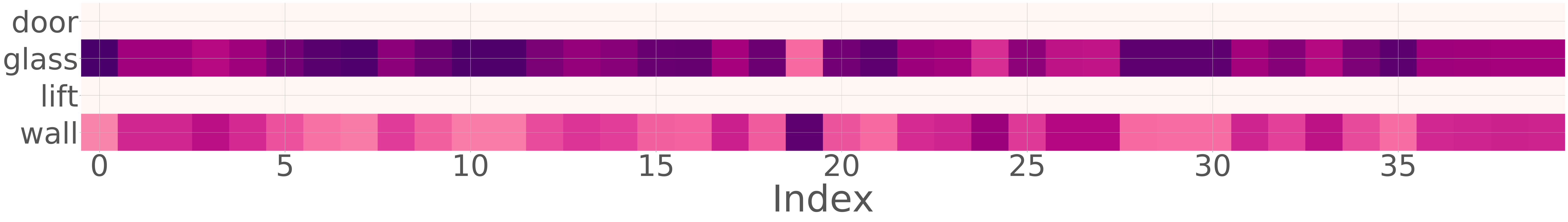} 
		\caption{Out of set: tables}
	\end{subfigure}%

	\begin{subfigure}[b]{0.47\textwidth}\centering
		\includegraphics[width=\columnwidth]{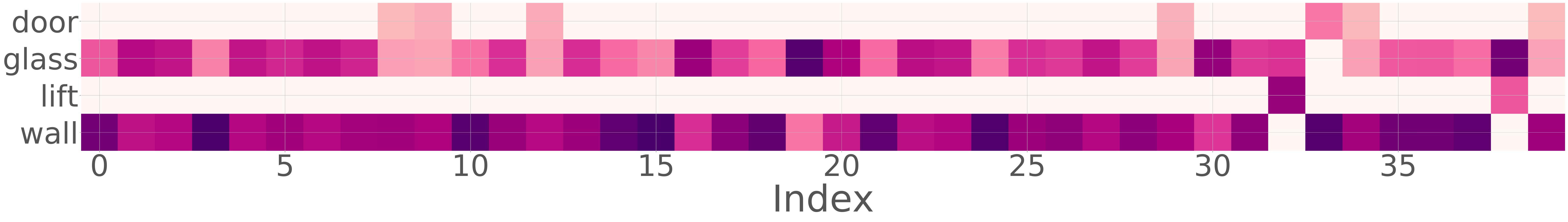} 
		\caption{Out of set: chairs}
	\end{subfigure}%

	\begin{subfigure}[b]{0.47\textwidth}\centering
		\includegraphics[width=\columnwidth]{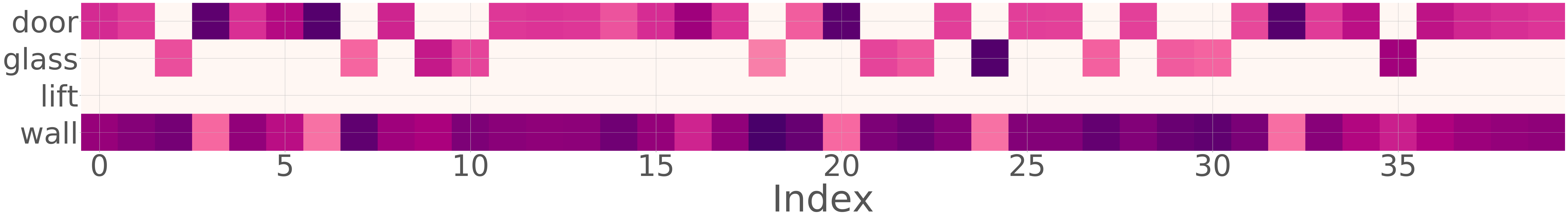} 
		\caption{Out of set: basins}
	\end{subfigure}%
\caption{Softmax distribution comparison between known classes and out-of-set classes. A dark color represents a large value i.e. high confidence and the horizontal axis denotes sample index. For known labels (top row), the distribution is unimodal. In contrast, the distribution of out-of-set samples are spread over multiple classes, yielding low classification scores.}
\label{fig:alien_objects}
\vspace{-0.4cm}
\end{figure}

As shown in Fig.~\ref{fig:alien_objects}, compared to the samples with the known labels, the probability distribution output by the softmax layer for three out-of-set objects are substantially more scattered and flat. Their resulting classification score is accordingly lower than the known samples. Based on $500$ samples from $5$ different alien objects (e.g., basins, tables, chairs, sofa and fridges.), we empirically found that a threshold of $0.92$ on the softmax classification score can correctly detect over $96\%$ of samples as unknowns. In the meantime, it only results in less than $2.2\%$ false negative rate for known samples. 

\subsection{System Efficiency} % (fold)
\label{sub:system_efficiency}

In the last experiment, we investigate the runtime latency, summarized in Tab.~\ref{tab:running_time}. Four platforms fitting the payload of mobile robots are used in our evaluation, including Raspberry Pi 3 (RPi 3), Raspberry Pi 4 (RPi 4), NVIDIA Jetson TX2 (TX2) and a mini netbook. In the implementations, we use TensorFlow Lite \cite{alsing2018mobile} to compress our models as per the convention of efficient on-device inference of DNNs. 
Tab.~\ref{tab:running_time} suggests that both map reconstruction and semantic mapping modules are able to run in real-time on all platforms. Even for the most challenging case (i.e., map reconstruction on RPi 3), a runtime of $2.58$s is also acceptable, because an input patch to our reconstruction network is generated by a robot crossing over a $6 \times 6 m^2$ square (see Sec.~\ref{sub:network_input}) while most ground robots' max speeds are $\leq$ 1m/s.

% In the last experiment, we investigate the system efficiency summarized in Tab.~\ref{tab:running_time}.  It is worth noting that under emergency situations (e.g., fire fighting) where timing is critical, latency is a dominant factor. To this end, we examine the runtime of grid map reconstruction and semantic mapping modules on four platforms that can match the robot payloads, including: Raspberry Pi 3 (RPi 3), Raspberry Pi 3 (RPi 4), NVIDIA Jetson TX2 (TX2) and a mini netbook that comes with the package in turtlebot 2. In implementation, we use TensorFlow Lite \cite{alsing2018mobile} to compress the models of the above two modules, which is a well known solution for efficient on-device inference of DNNs. 
% As we can see in Tab.~\ref{tab:running_time}, on low-end devices such as RPi 3 and RPi 4, the map reconstruction time takes 1 $\sim$ 2 seconds per patch. Recall that a patch is generated by a robot crossing over a $6 \times 6 m^2$ square, the reconstruction time is able to match the speeds of mobile robots online. Remarkably, when more computation power are allowed, such as on TX2 and netbook, the reconstruction time can be substantially shortened to sub-seconds. On the other side, owning to the lightweight structure, the semantic mapping takes sub-millisecond to process a scan/frame even the most resource-constrained platform. As the sampling rate of our mmWave radar is between $10 \sim 30$ frames/s, such efficiency can enable fast semantic mapping in real time.

%!TEX root = ../main.tex
\section{Related work}
\label{sec:relatedwork}

\noindent \textbf{RF-based Imaging and Tracking.}
Signal reflection of RF waves has been widely leveraged to perform imaging and target tracking. In the WiFi bands, researchers have used commodity WiFi chips \cite{depatla2015x,huang2014feasibility,pu2013whole,liu2012push,sun2015widraw,jiang2013hallway} to imagine static objects, localize humans and recognize predefined hand gestures. Additionally, by leveraging a specialized FMCW radar \cite{adib2013see,adib2015capturing,adib2015multi,adib20143d,zhao2018rf,zhao2018through,zhao2019through}, WiFi signals can be used to accurately track/imagine human body dynamics, as well as recover pose estimation under NLOS scenarios. 
In the vein of mmWave-based tracking, Babak et al. use FMCW hardware and apply SAR with sparse measurements in absence of device movement noises \cite{mamandipoor201460}, while Xu et al. uses customized mmWave probe to recover human speeches via throat localization \cite{xu2019waveear}. On the side of environment sensing, research effort has been devoted to pinpoint indoor major reflectors, thereby combating the environment sensitivity of mmWave communications \cite{palacios2017jade,wei2017facilitating,zhou2019autonomous}. Nevertheless, major reflectors are still sparse points which are incomparable to the dense grid maps to first responders. Recent works \cite{zhu201560ghz,zhu2017object} pioneered the research of low-cost mmWave devices to explicitly image objects. By continuously moving or navigating in front of a specific object, they can infer the geometry of small indoor objects. However, such iterative mapping and navigation strategy violates limited time budgetsin search and rescue scenarios. 
In contrast, \sysname uses a low-cost off-the-shelf mmWave radar to reconstruct a dense occupancy grid map while a robot travels freely in an environment. 

% or a navigation method that is sensitive to multi-path reflections with large imaging latency. 

\noindent \textbf{RF-based Material/Object Identification.} 
By charactering the reflection intensity of RF signals, the RSA system \cite{zhu2015reusing} measures the reflected mmWave signals at multiple locations and then use an aggregated value to identify a target's surface material. A similar work is RadarCat \cite{yeo2016radarcat}, a contact based material identification systems leveraging 60 GHz signals. \sysname differs from the RSA and Radar in that it does not require multiple measurements at different locations nor a physical contact with the target material. Recent studies also found mmWave signals can detect and classify hidden electronic devices \cite{li2018eye} and even screen activities \cite{li2020wavespy}. On the other side, WiFi CSI \cite{feng2019wimi}, UWB \cite{dhekne2018liquid} and RFID \cite{wang2017tagscan} have recently been utilized to identify materials based on their phase and RSS readings. However, these systems are sensitive to the calibrated positions of pairs of transmitters and receivers, while \sysname is a single-chip solution to mobile robotic platform.  

\noindent \textbf{Indoor Mapping/Imaging with non-RF Sensors.} Optical sensors, such as RGB cameras \cite{gao2014jigsaw,dong2015imoon}, laser rangers \cite{surmann2003autonomous} and stereo cameras \cite{henry2014rgb} are established sensor modalities to produce accurate indoor maps. However, these sensors are notoriously fragile under adverse vision conditions, e.g., darkness, glare and smoke debris. Acoustic sensors such as microphones \cite{mao2018aim,pradhan2018smartphone,zhou2017batmapper} are recently found to be effective for indoor mapping and object imaging but their performances are restricted by limited sensing ranges and sensitive to environmental noises as well as sound-absorbing materials.

\begin{table}[!t]
	\centering
	\small
	\caption{Runtime efficiency of key modules in \sysname.}
	\label{tab:running_time}
	\begin{tabular}{|c|c|c|c|c|}
	\hline
	 & \textbf{RPi 3} & \textbf{RPi 4} & \textbf{TX2} & \textbf{Netbook} \\ \hline
	\textbf{Map Recon. (s)} & 2.58 & 1.01 & 0.65 & 0.33 \\ \hline
	\textbf{Semantic Mapping (ms)} & 0.17 & 0.08 & 0.06 & 0.02 \\ \hline
	\end{tabular}
	\vspace{-0.4cm}
\end{table}

%!TEX root = ../main.tex
\section{Limitations and Future Work}
\label{sec:limitations}

This work focuses on a proof-of-principle mapping using mmWave radar, towards our vision of augmenting emergency response with low-cost mobile sensing systems. There are limitations and a number of avenues for future exploration. Firstly, the turtlebot platform is not rugged enough for a real disaster situation. Other more robust platforms have been designed to tackle this problem \cite{delmerico2019current}, e.g. the use of tracked or snake-like robots. Aerial micro-robots are also a potential alternative for rapid exploration, and the form-factor of the single chip radar is ideally suited as a primary sensor for these agents. Secondly, further trials need to be performed under diverse conditions such as different buildings, varying obscurants (e.g. dust in a factory) and under real emergency conditions. Thirdly, we have focussed on using a single agent to build a map, in future work we will explore how to use swarms of robots to cooperatively explore and build the map e.g. by using SLAM~\cite{achtelik2012sfly}.

%!TEX root = ../main.tex

\section{Conclusions}
\label{sec:conclusions}

Indoor mapping in low-visibility environments full of airborne particulates is a challenging yet important problem. Particularly of importance to emergency responders, 
an accurate map can significantly aid in situational awareness and become a life saver in search and rescue scenarios. To this end, \sysname used a mmWave radar on a mobile robot to create a dense map that indicates place reachability and object semantics on the map. We also demonstrated how another agent could relocalize within the map. With extensive experiments in different indoor environments and under smoke-filled conditions, we show the performance of reconstruction, semantic classification and system efficiency of \sysname, demonstrating its ability to generalise to previously unseen environments.

% We presented \sysname, a learning-based inductive method for obtaining dense occupancy grid maps from low-cost mmWave radar sensors, using cross-modal supervision from partial labels from a lidar. By leveraging the structure of indoor scenarios, the model is able to reconstruct the shape of novel environments and, to some extent, cope with noisy odometry and smoke-filled scenarios. The limitation of the approach lies in the potential inaccuracy of labels (e.g., in presence of glass and reflective materials for lidar). Future work will be devoted to automatically detect such materials from the raw mmWave measurements, that are robust to presence of glass and metal.

\begin{acks}
We thank all anonymous reviewers and our shepherd for their helpful comments. This work was supported, in part, by the awards 70NANB17H185 and 60NANB17D16 from the U.S. Department of Commerce, National Institute of Standards and Technology (NIST) and the UK EPSRC through Programme Grant EP/M019918/1.
\end{acks} 

% \clearpage 
\balance
\bibliographystyle{ACM-Reference-Format}
\bibliography{ref}

\end{document}